\documentclass[titlepage,pra,aps,final,%
nofootinbib]{revtex4}
\usepackage{graphics}
\usepackage{graphicx,colordvi}
\usepackage{lscape,multirow}
\usepackage{amsmath}

\expandafter\ifx\csname package@font\endcsname\relax\else
\expandafter\expandafter
\expandafter\usepackage
\expandafter{\csname package@font\endcsname}
\fi
\DeclareRobustCommand\substyle{\name@idx{document substyle}}
\DeclareRobustCommand\classoption{\name@idx{document class option}}
\DeclareRobustCommand\classname{\name@idx{document class}}
\def\name@idx#1#2{{\ttfamily#2}
\index{#2\space#1=\string\ttt{#2}\space#1}\index{#1>#2=\string\ttt{#2}}}

\newcommand{\beq}[0]{\begin{equation}}
\newcommand{\eeq}[0]{\end{equation}}
\newcommand{\bea}[0]{\begin{eqnarray}}
\newcommand{\eea}[0]{\end{eqnarray}}
\newcommand{\minitab}[2][l]{\begin{tabular}{#1}#2\end{tabular}}
\usepackage{amssymb}

\def\d{\hbox{d}}
\def\be{\begin{equation}}
\def\ee{\end{equation}}
\def\bea{\begin{eqnarray}}
\def\eea{\end{eqnarray}}
\def\l{\label}

\def\hahat{\hat{H}}

\def\hahat0{\hat{H}_0}

\def\cos{\hbox{cos}}

\def\cosh{\hbox{cosh}}
\def\sinh{\hbox{sinh}}

\def\exp{\hbox{exp}}

\def\sinh{\hbox{sinh}}

\def\vareps{\varepsilon}

\def\siml{\hbox{\kern.1em \lower.6ex \hbox{$\sim$} \kern-1.12em
          \raise.6ex \hbox{$<$} \kern.1em }}
\def\simg{\hbox{\kern.1em \lower.6ex \hbox{$\sim$} \kern-1.12em
          \raise.6ex \hbox{$>$} \kern.1em }}

\def\siml{\hbox{\kern.1em \lower.6ex \hbox{$\sim$} \kern-1.12em
 \raise.6ex \hbox{$<$} \kern.1em}}
\def\simg{\hbox{\kern.1em \lower.6ex \hbox{$\sim$} \kern-1.12em
 \raise.6ex \hbox{$>$} \kern.1em}}

\newcommand{\beqar}{\begin{eqnarray}}
\newcommand{\eeqar}[1]{\label{#1} \end{eqnarray}}

\begin{document}

\title{Microscopic-macroscopic level
  densities for low excitation energies
}

\author{ A.G.~Magner$^{a,b}$,
A.I.~Sanzhur$^{a}$, S.N.~Fedotkin$^{a}$,
A.I.~Levon$^{a}$, U.V.~Grygoriev$^{a}$, S.~Shlomo$^{b}$\\[2ex]
{\em $^{a}$Institute for Nuclear Research NASU, 03028 Kyiv, Prospekt Nauky 47, Ukraine}\\
{\em $^{b}$Cyclotron Institute, Texas A\&M University, College Station, Texas 77843, 120 Spence Str., USA}}

\begin{abstract}
  Level density $\rho(E,{\bf Q})$ is derived within the 
  micro-macroscopic approximation (MMA) for a 
  system of  strongly  interacting Fermi particles with the energy $E$ and
  additional
  integrals of motion ${\bf Q}$,
  in line with several topics of the universal and fruitful
  activity of A.S. Davydov.
  Within the extended Thomas Fermi and semiclassical periodic orbit
  theory 
  beyond the Fermi-gas saddle-point method we
  obtain $~~\rho \propto I_\nu(S)/S^\nu$,~~ where $I_\nu(S)$ is the
  modified Bessel function
  of the entropy $S$. For small shell-structure contribution
  one finds
  $\nu=\kappa/2+1$,  where $\kappa$  is
  the number of additional integrals of motion.
  This integer number is
  a dimension of ${\bf Q}$, ${\bf Q}=\{N, Z, ...\}$ for the case of
  two-component atomic nuclei, where
  $N$ and $Z$ are the numbers of neutron and protons, respectively.
   For much larger shell structure 
    contributions,
one obtains,
$\nu=\kappa/2+2$.
      The MMA level density $\rho$ 
  reaches 
  the well-known 
    Fermi gas asymptote 
  for 
  large excitation energies,
  and the finite micro-canonical combinatoric
  limit for
  low excitation energies. The additional integrals of motion can be also the
  projection of the
  angular momentum of a nuclear system for nuclear rotations of deformed nuclei,
  number of excitons for collective dynamics, and so on. 
  Fitting the MMA total level density, $\rho(E,{\bf Q})$,
  for a set of the integrals of motion
  ${\bf Q}=\{N, Z\}$, to 
    experimental data on a long nuclear isotope chain
    for
  low excitation energies,
  one obtains the results for the 
    inverse level-density parameter $K$,
    which differs 
    significantly from  those of
      neutron resonances, due to shell, isotopic asymmetry,
and pairing effects.
     
\end{abstract}

\maketitle
%





\section{Introduction}\label{sec1}
The statistical level density is a fundamental tool for the description of 
many
properties of finite Fermi systems; see, e.g.,
Refs.~\cite{Be36,Er60,GC65,BM67,LLv5,Ig83,So90,Sh92,Ig98,Ju98,AB00,AB03,OA03,%
AB15,EB09,Gr13,ZS16,AB16,KZ16,HJ16,KS18,ZK18,ZH19,Gr19,KS20,KZ20,GA21,FA21}
 for atomic nuclei.
Usually, the level density, 
$\rho(E,{\bf Q})$, for a nuclear system is defined as function of
its energy, $E$,
and a number of the additional integrals of motion, ${\bf Q}$.
These integrals of motion can be specified, for instance, as
${\bf Q}=\{N, Z, M\}$,
where $N$, $Z$, and $M$ 
are the neutron, proton numbers, and projection of the angular momentum
on a laboratory-fixed
coordinate system,
respectively. The level density can be presented as
the inverse Laplace transformation of the partition function
$\mathcal{Z}(\beta,\boldsymbol\alpha)$,
where $\beta$ and
$\boldsymbol\alpha=\{\alpha^{}_{N},\alpha^{}_{Z}$, $\alpha^{}_{M}\}$ are
Lagrange multipliers arguments of the partition function $\mathcal{Z}$.
The Lagrange multipliers $\alpha^{}_{N}$ and $\alpha^{}_{Z}$ are determined
by the conservation of neutron
$N$ and proton $Z$ numbers, respectively, and
$\alpha^{}_{M}$ provides the conservation of the projection $M$ of the
angular momentum through the corresponding saddle point condition.
In this respect, we pay attention also to developments of
Refs.~\cite{Bj74,BM75,Ju98,OA03,Gr13,Gr19}
based on Bohr \& Mottelson \cite{BM75} and Davydov with co-workers
\cite{DF58,DC60} theory
of the nuclear axially-symmetric and non-axially-symmetric rotations,
respectively.
Within the  grand
canonical ensemble, 
one can apply the standard Darwin-Fowler method
for the saddle-point 
integration
over all variables,
including 
$\beta$, which is related to the total energy $E$; see Refs.~\cite{Er60,BM67}.
This method assumes a large
excitation energy $U$, so  that the temperature $T$ is
related to a  well-determined
saddle point
in the inverse Laplace integration variable
$\beta$ for a finite Fermi 
    system of large particle numbers and angular momenta. 
However,
many experimental data 
also 
exist for a low-lying part of the
excitation energy $U$,
where  such a saddle point does not exist; see, e.g., Ref.~\cite{Le20}.
Therefore, the integral over the Lagrange multiplier
$\beta$ in the inverse Laplace
  transformation of the partition function
$\mathcal{Z}(\beta,\boldsymbol\alpha)$  
  should be carried out  more 
  accurately  beyond the standard saddle-point method, 
  see
 Refs.~\cite{KM79,NPA,PRC,IJMPE}. 
 For other variables
  related 
  to the  neutron $N$ and proton $Z$ numbers, and projection $M$
of the angular momentum $I$, one can
apply the saddle point method assuming that
  $N$, $Z$, and $I$ are large. For the critical points in these saddle-point integrations
  if high-order second-derivatives of entropy over the corresponding Lagrange multiplier,
  $\boldsymbol\alpha$, are zeros or infinities. We will consider the results
  for level density and fluctuations near such catastrophe points
  in a forthcoming separate work. But, in the following we will consider
  a critical point at the
  zero-excitation energy limit
  ($\beta \rightarrow \infty$) where all high-order derivatives of the entropy over $\beta$
  are zeros, as suggested in Refs.~\cite{KM79,NPA,PRC,IJMPE}. This catastrophe point is
  similar to that in the zero-deformation limit for particles in a mean field for solving the
  symmetry breaking phenomenon within the semiclassical
  periodic-orbit theory (POT)
  \cite{SM76,SM77,BB03,MA02,MA06,MK16,MY11,MA17}.
  As shown in Refs.~\cite{KM79,PRC,IJMPE}, the level density calculations
  for large particle
    numbers and angular momenta, and for a deeper understanding of the 
correspondence between the classical and the quantum approach,
it is worthwhile to analyze the shell 
effects in the level density $\rho$
 (see Refs.~\cite{Ig83,So90}) 
and in the entropy, $S$, using 
 the semiclassical POT \cite{SM76,SM77,BB03,MY11}.
This theory, based on the semiclassical
time-dependent propagator, allows
obtaining the total level density, energy,
canonical free energy, and grand canonical 
potential, in terms of the smooth extended Thomas-Fermi 
term and periodic orbit
correction taking into account the symmetry breaking phenomena.
  Notice also that 
  other semi-analytical methods were suggested in the literature
  \cite{JB75,BJ76,PG07}
  to overcome divergence
  of the full saddle-point method 
  for the low
excitation-energy limit, $U \rightarrow 0$.  

 More  general microscopic formulation of the energy level density 
 for mesoscopic systems, in particular for nuclei,  which removes
 the singularity at small excitation energies, is discussed 
in Ref.~\cite{ZH19}, see also references therein.  
One of the microscopic ways for accounting
for interparticle interactions beyond the mean field (shell model)
     in the level density calculations
     was suggested within the Monte-Carlo Shell Model \cite{Or97,AB03,AB15}.
     Another successful 
     approach for taking into
     account the interparticle
         interactions above the simple shell model
      is given by
     the moments 
     method \cite{KZ16,Ze16,ZK18,KZ20,Ze96}. The main ideas are based on
     the random matrix theory, see Refs.~\cite{Po65,Ze96,Ze16,Me04}.

In a semiclassical formulation of a 
unified microscopic 
canonical 
and macroscopic grand-canonical 
approximation (shortly, MMA) for the level density, we
derived a simple nonsingular analytical
expression of the
level density $\rho$ for neutron-proton asymmetric nuclei \cite{IJMPE}.
The MMA approach satisfies
the two well-known limits. 
One of them is the 
Fermi gas asymptote, 
$\rho \propto \exp(S)$,  
for a large entropy $S$.
    The opposite limit for small $S$, or excitation energy $U$, 
is the  combinatorics expansion \cite{St58,Er60,Ig72} in powers of $S^2$. 
 For small excitation energies, the empiric formula,
    $\rho\propto \exp[(U-E_0)/T]$, with free parameters $E_0$,  $T$, and
  a  pre-exponent factor,
  was suggested for the description of the level density
  of the excited low
  energy states in Ref.~\cite{GC65}.
Later, this formula was named a constant
``temperature'' model (CTM), 
see also Refs.~\cite{ZK18,ZH19,KZ20}.
The ``temperature'' was considered as an
``effective temperature'' which is
related to the excitation energy because 
of no direct physical meaning
of temperature for low energy states exists.
Following the development of
    Refs.~\cite{NPA,PRC} we will show below that 
the MMA
has the same power  expansion as the constant
``temperature'' model 
for low energy states
at small excitation energies $U$.

Such an MMA for  the
level density $\rho$ was 
suggested in Ref.~\cite{KM79},
 within 
the Strutinsky shell correction method \cite{St67,BD72,BK72} 
based on the Landau-Migdal
quasiparticle theory named as the Finite Fermi System
Theory \cite{La58,AK59,MI67,HS82}.
A mean field potential is used for
calculations of the
energy shell corrections, $\delta E$. The total nuclear energy, $E$,  is the
sum
of these corrections and smooth macroscopic liquid-drop component \cite{MS69}
which can be well approximated by the extended Thomas-Fermi approach
\cite{BG85,BB03}. Thus, 
    within the semiclassical approximation to the
    Strutinsky shell correction method,
the interactions between particles, averaged
over particle numbers, 
    i.e., over many-body microscopic quantum states 
    in realistic
nuclei, 
are approximately taken into
account through the  
extended Thomas-Fermi component beyond the mean field.
Neglecting small {\it residual}-interaction corrections (see Ref.~\cite{IJMPE}) 
    beyond  
     the macroscopic extended Thomas-Fermi  approach and Strutinsky's
        shell corrections,
one can present \cite{KM79} 
the level density $\rho$  
in terms of the
modified Bessel function
of the entropy variable in the case of
small
 thermal excitation energy $U$ as compared to the
 rotational energy.

The MMA
approach \cite{KM79}
was extended \cite{NPA,PRC,IJMPE}
for the description of shell, rotational, pairing and isotopic asymmetry effects 
on the level density 
itself for
larger excitation energies $U$ in 
nuclear systems.
We will apply in the following
this MMA for
analytical level-density calculations for other nuclear systems with larger deformations
and angular momenta.
The level
density parameter $a$ and moment of inertia $\Theta$
 for asymmetric neutron-proton nuclear systems at high spins
are key quantities under
intensive experimental and theoretical 
 investigations \cite{Er60,GC65,BM67,Ig83,Sh92,So90,EB09,ZH19,KS20}. 
 As mean values of the level density parameter
 $a$ are largely
proportional to
the total particle number $A=N+Z$,
the inverse level-density parameter,
$K=A/a$, is conveniently introduced
to exclude 
a basic mean particle-number dependence in $a$.
Smooth properties of this 
function of the nucleon number $A$
have been studied
within the framework of the
self-consistent extended Thomas-Fermi 
approach \cite{Sh92,KS18}, see also the
study of shell effects in
 one- and two-component nucleon systems in Refs.~\cite{NPA,PRC,IJMPE}.
 However,  
the statistical
level density for neutron-proton asymmetric rotating nuclei
is still 
an attractive subject. For instance, 
within the Strutinsky's shell correction approach \cite{BD72},
the major shell
effects in the distribution of single-particle (quasiparticle) 
states near
the Fermi surface 
are quite different for neutrons and protons of asymmetric nuclei, 
especially for nuclei far from the $\beta$-stability line. The shell effects in the level
density is expected to be important also for nuclear fission \cite{BD72}.
Another interesting subject is the
influence of the shell effects on the the moment of inertia
at high spins \cite{RB80,St87} and thereby on the level density.
In thhe present work we concentrate on
low energy states of
nuclear excitation-energy spectra  below 
the neutron resonances 
for large chains of the nuclear deformed 
isotopes.

The structure of the paper is the following.
The level density $\rho(E,{\bf Q})$ is derived
within the MMA 
by using the semiclassical periodic-orbit theory 
in Sec. \ref{sec2}.
The general shell, isotopic asymmetry, and rotation
effects (Subsection~\ref{subsec2-1})
are first discussed within the standard saddle-point method
 asymptote (Subsection~\ref{subsec2-2}).
We then extend 
the standard 
saddle-point method to a more general MMA approach for describing
the analytical transition from large to small  
excitation energies $U$,
taking 
essentially into account the shell and isotopic
asymmetry effects (Subsection~\ref{subsec2-3}). The
level density of rotating systems are presented in Sec.~\ref{sec3}.  
In Section \ref{sec4}, we compare our analytical  MMA results for the level
density $\rho$, and the inverse level-density
    parameter $K$,
with experimental
data for a large isotope chain 
as typical examples of heavy isotopically 
asymmetric and deformed nuclei. Our results will be summarized in Section
\ref{sec5}.
Some details of 
the POT,
and of the 
    standard saddle-point method are presented in  Appendixes
A and B, respectively.

\section{Microscopic-macroscopic approach }\label{sec2}

\subsection{General points}\label{subsec2-1}

 For 
a statistical
description of the level density of a
Fermi system in
  terms of the conservation  variables; 
the total energy, $E$, and additional integrals of motion ${\bf Q}$; 
one
can begin with
the micro-canonical expression for the level density,
\bea\l{dendef1}
&\rho(E,{\bf Q})=
\sum\limits_i\!\delta(E-E_i)~\delta({\bf Q}-{\bf Q}_i) \nonumber\\ 
&\equiv
\int \frac{d \beta d \boldsymbol\alpha}{(2\pi i)^{\kappa+1}}~
  \exp\left[S(\beta,\boldsymbol\alpha)\right]~.
\eea
Here, $E_i$ and ${\bf Q}_i$
represent the system spectrum of the $\kappa+1$ degree of freedom,
where $\kappa$ is the
number of integrals of motion other than the energy $E$.
For instance, for a nucleus
one has
${\bf Q}_i=\{N_i,Z_i,M_i\}$ ($\kappa=3$), where $N_i$ and $Z_i$
are the number
of neutrons and protons, respectively, and $M_i$
is the projection of angular
momentum ${\bf I}_i$ to
  a laboratory fixed-axis system. We assume that there are
    no external forces acting on the nucleus.
The entropy $S$ is determined by the partition function
$\mathcal{Z}(\beta,\boldsymbol\alpha)$,
\bea\l{entnp}
&S(\beta,\boldsymbol\alpha)=\ln \mathcal{Z}(\beta,\boldsymbol\alpha)
+\beta E -\boldsymbol\alpha {\bf Q}\nonumber\\ 
&=\beta\left(E-\Omega- \boldsymbol\lambda{\bf Q}\right)~, 
\eea
where $\boldsymbol\alpha=\boldsymbol\lambda\beta$, $\boldsymbol\lambda=
\{\lambda_n,\lambda_p,\hbar \omega\}$ with the neutron chemical potentials $\lambda_\tau$
and $\tau=\{n,p\}$ 
being the isotope subscript for a nucleus. The Lagrange multipliers $\lambda_\tau$ provide
the conservation of the neutron, $N_i$, and proton, $Z_i$, numbers in a nucleus.
We introduced also a frequency $\omega$
of rotations around
the axis of a space-fixed coordinate system as another Lagrange multiplier which corresponds to
the conservation of the angular momentum projection $M_i$.
The entropy $S$, partition $\mathcal{Z}$, and
potential $\Omega$ functions 
are considered 
for arbitrary values of arguments $\beta$ and  $\boldsymbol\alpha$, and
$\Omega=-\ln \mathcal{Z}/\beta$~. 
The integral on the right-hand side 
of Eq.~(\ref{dendef1}) is the
standard inverse Laplace transformation of the partition function $\mathcal{Z}$.
For large excitation energies,
when the saddle points of the integrals
in Eq.~(\ref{dendef1}) over all
variables $\beta$ and $\boldsymbol\alpha$
exist \cite{Ig83,So90},
we have the standard 
    entropy $S$, partition function $\mathcal{Z}$ and thermodynamic
potential $\Omega$ \cite{LLv5}. 
 In the axially-symmetric mean field of the Strutinsky's shell correction method \cite{BD72},  
 the single-particle (quasiparticle) 
 level density, $g(\varepsilon,m)$, where $\varepsilon$ and $m$ are the single-particle energies
 and projection of the angular momentum on any axis of a space-fixed coordinate system,
 can be 
 written \cite{BM67} as a sum of the neutron and proton components in a nucleus,
$g=g_n+g_p$~. This leads to a similar isotopic decomposition for the 
potential $\Omega$, 
$\Omega=\Omega_n+\Omega_p$. For axially symmetric nucleus the potential
$\Omega_\tau$ is given by [see Eq.~(\ref{parfun}) for the
      partition function  $\mathcal{Z}$] 
\be\l{OmFnp}
\Omega_\tau\approx -
\int\limits_0^\infty\frac{d\varepsilon d m}{\beta}
g_\tau(\varepsilon,m)
\ln\left\{1+\exp\left[\beta\left(\lambda_\tau-
  \varepsilon -\hbar \omega m\right)\right]\right\}.
\ee
The 
level density, $g_{\tau}(\varepsilon,m)$, within
the Strutinsky's shell-correction method \cite{BD72}, 
is a sum of
the statistically averaged smooth, $\tilde{g}_\tau(\varepsilon,m)$, component,
and the oscillating shell component,
$\delta g_{\tau}(\varepsilon,m)$, correction
    for an arbitrary axially-deformed nucleus, slightly averaged over the 
single-particle energies,
\be\l{gdecomp}
g_\tau(\varepsilon,m)\cong \tilde{g}_\tau(\varepsilon,m)+
\delta g_\tau(\varepsilon,m)~. 
\ee
Within the semiclassical POT
\cite{SM76,SM77,BB03,PRC,MK78} (Appendix \ref{appA}),
the smooth and
oscillating parts
of the 
level density, $g_\tau(\varepsilon,m)$, Eq.~(\ref{gdecomp}),
can be approximated, with good accuracy, by
the extended Thomas-Fermi 
level density,
$\tilde{g}_\tau \approx g^{(\tau)}_{\rm \tt{ETF}}$,
and the
periodic-orbit contribution, $\delta g_\tau\approx
\delta g^{(\tau)}_{\rm scl}$, respectively, e.g. see 
Eq.~(\ref{goscemsph}) for spherically symmetric potentials.
Using the POT
  decomposition, Eqs.~(\ref{gdecomp}) and (\ref{OmFnp}),
one finds $\Omega_\tau=\tilde{\Omega}_\tau+\delta \Omega_\tau$.
For a smooth (ETF) part of this $\tau$-potential,
$\tilde{\Omega}_\tau\approx \Omega^{(\tau)}_{\rm\tt{ETF}}$, one can use 
 the result \cite{KM79,KS20}:
\bea\l{TFpotF}
&\tilde{\Omega}_\tau
=\tilde{E}_\tau 
-\lambda_\tau \mathcal{N}_\tau
-\frac{\pi^2}{6\beta^2}\tilde{g}_\tau\nonumber\\
&-\frac{1}{2}\tilde{\Theta}_\tau(\lambda_\tau) \omega^2~,\quad
\mathcal{N}_\tau=\{N,Z\}~.
\eea
 Here, $\tilde{E}_\tau\approx E^{(\tau)}_{\rm\tt{ETF}}$ is the
 nuclear extended Thomas-Fermi 
 energy 
component (or the corresponding liquid-drop
energy), and
$\lambda_\tau$
is approximately the smooth
chemical potential for neutron ($n$) and proton ($p$) subsystems in the
shell correction method \cite{BD72}.
The moment of inertia, $\Theta=\Theta_n+\Theta_p$,
is decomposed in terms of a smooth (ETF) part, $\tilde{\Theta}_\tau=\Theta_{\rm ETF}$,
and shell correction $\delta \Theta_\tau$,
$\Theta_\tau=\tilde{\Theta}_\tau+\delta \Theta_\tau$. 
With the help of the POT 
\cite{SM76,SM77,BB03,PRC,MK78}, one
obtains \cite{KM79} for the oscillating (shell)
component, $\delta \Omega_\tau$,  Eq.~(\ref{OmFnp}),
\bea\l{potoscparFnp}
&\delta \Omega_\tau= - 
\int\limits_0^\infty\frac{ d\varepsilon d m}{\beta}
\delta g_\tau(\varepsilon,m)\nonumber\\
&\times\ln\left\{1+\exp\left[\beta\left(\lambda_\tau-
  \varepsilon-\hbar \omega m\right)\right]\right\}
\cong
\delta \Omega^{(\tau)}_{\rm scl}
=\delta F^{(\tau)}_{\rm scl}\ .
\eea
%
One can find the explicit POT
expressions  for the
semiclassical free-energy shell correction,
$\delta F^{(\tau)}_{\rm scl}$,
or $\delta \Omega^{(\tau)}_{\rm scl}$, in the corresponding variables,
within the spherical
mean-field approximation in Refs.~\cite{KM79,PRC}.
For nonrotating but more general deformed nuclei
we incorporate 
the following 
explicit periodic-orbit (PO) expression \cite{KM79,BB03}: 
\be\l{FESCFnp}
\delta F^{(\tau)}_{\rm scl} \cong \sum^{}_{\rm PO} F^{(\tau)}_{\rm PO}~,
\ee
where
\bea\l{dFESCFnp}
&F^{(\tau)}_{\rm PO}= E^{(\tau)}_{\rm PO}~
\frac{x^{(\tau)}_{\rm PO}}{
  \sinh\left(x^{(\tau)}_{\rm PO}\right)}~,\\
&x^{(\tau)}_{\rm PO}=
\frac{\pi t^{(\tau)}_{\rm PO}}{\hbar \beta}~.\l{x}
\eea
%
The periodic-orbit component $E^{(\tau)}_{\rm PO}$ of the semiclassical 
shell-correction energy was derived earlier in
Ref.~\cite{SM76} to be,
\be\l{dEPO0Fnp}
 \delta E^{(\tau)}_{\rm scl}=
\sum^{}_{\rm PO}E^{(\tau)}_{\rm PO}
=\sum^{}_{\rm PO}\frac{\hbar^2}{(t^{(\tau)}_{\rm PO})^2}\,
g^{(\tau)}_{\rm PO}(\lambda_\tau)~.
\ee
where $g^{(\tau)}_{\rm PO}(\lambda_\tau)$ is the PO component of the total single-particle
level density $g_\tau(\lambda_\tau)=\sum_mg_\tau(\lambda_\tau,m)$ (Appendix \ref{appA}).
Here, $t^{(\tau)}_{\rm PO} = 
\Upsilon t^{\Upsilon=1}_{\rm PO}(\lambda_\tau)$
is the period 
of particle motion
along 
a periodic  orbit 
(taking into account its repetition, or period number 
$\Upsilon$), and $t^{\Upsilon=1}_{\rm PO}(\lambda_\tau)$ is the
period of the neutron ($n$) or proton ($p$) motion along the
primitive ($\Upsilon=1$)
periodic orbit in the corresponding $\tau$ potential well 
    with the same
radius, $R=r^{}_0A^{1/3}$.
The period $t^{(\tau)}_{\rm PO}$ 
(and $t^{\Upsilon=1}_{\rm PO}$), and
the partial oscillating level density component, $g^{(\tau)}_{\rm PO}$,
are taken at the chemical potential, $\varepsilon = \lambda_\tau$,
see also Eq.~(\ref{goscsem}) for the semiclassical 
level-density shell correction
(Appendix \ref{appA} and Refs.~\cite{SM76,BB03}). The semiclassical expressions,
Eqs.~(\ref{TFpotF}) and (\ref{potoscparFnp}), are valid for a large
relative action, $\mathcal{S}^{(\tau)}_{\rm PO}/\hbar \sim A^{1/3} \gg 1$~.

Then, expanding 
$x^{(\tau)}_{\rm PO}/\sinh(x^{(\tau)}_{\rm PO})$,
Eq.~
(\ref{x}), in the shell correction $\delta \Omega_\tau$
[Eqs.~(\ref{potoscparFnp}) and (\ref{FESCFnp})]
in powers of  $1/\beta^2$
up to the quadratic terms, $\propto 1/\beta^2$,
one obtains for an adiabatic rotation, 
\be\l{OmadFnp}
\Omega_\tau \approx E^{(\tau)}_0-\lambda_\tau
\mathcal{N}_\tau-\frac{a_\tau}{\beta^2} 
-\frac{1}{2}\Theta_\tau(\lambda_\tau) \omega^2~,
\ee
where $E^{(\tau)}_0$  is the neutron, or proton ground 
state energy, 
$E^{(\tau)}_0=\tilde{E}_\tau+\delta E_\tau$,
and $\delta E_\tau$ is the energy shell correction of the
corresponding cold system,
$\delta E_\tau \approx \delta E^{(\tau)}_{\rm scl}$
[see Eq.~(\ref{dEPO0Fnp}) and Appendix \ref{appA}].
In Eq.~(\ref{OmadFnp}),
$a_\tau$
is the level density parameter with  
a decomposition which is similar to Eq.~(\ref{gdecomp})
 at the $\tau $ chemical potential $\lambda_\tau$,
\be\l{denparnp}
a_\tau=\frac{\pi^2}{6}~g_\tau=\tilde{a}_\tau+\delta a_\tau~,
\ee
where $\tilde{a}_\tau$ is the extended Thomas-Fermi 
component and
$\delta a_\tau$ is the periodic-orbit
shell correction. Simple explicit expressions for the level density parameter,
Eq.~(\ref{denparnp}), with the $\omega$ dependence and its ETF and POT
components in the case of a spherical mean-field
is given in Ref.~\cite{PRC}. For the nonrotating case, one has
the following simple expressions:
\be\l{daFnp}
\tilde{a}_\tau 
\approx \frac{\pi^2}{6} g^{(\tau)}_{\rm \tt{ETF}}(\lambda_\tau), \quad
\delta a_\tau \approx 
\frac{\pi^2}{6}\delta g^{(\tau)}_{\rm scl}(\lambda_\tau)~.
\ee
For the extended Thomas-Fermi 
component \cite{BG85,BB03,KS18,KS20}, $g^{(\tau)}_{\rm \tt{ETF}}$,
one takes 
into account the
self-consistency with Skyrme forces \cite{AS05}. 
For the semiclassical PO
level-density shell corrections \cite{SM76,SM77,BB03,MY11,PRC},
$\delta g^{(\tau)}_{\rm scl}(\lambda_\tau)$, we use
Eq.~(\ref{goscsem}).

Expanding the entropy, Eq.~(\ref{entnp}), over the Lagrange multipliers
$\boldsymbol\alpha$ near the saddle point,
$\boldsymbol\alpha^\ast$, 
one can use the saddle point equations
(particle number and angular-momentum projection conservation equations),
\be\l{Seqsdnp}
\beta^{-1} \left(\frac{\partial S}{\partial \boldsymbol\lambda}\right)^\ast\equiv
-\left(\frac{\partial \Omega}{\partial \boldsymbol\lambda}\right)^\ast-{\bf Q}=0~,
\ee
where ${\bf Q}=\{N,Z,M\}$ and
$\boldsymbol\lambda=\{\lambda_n,\lambda_p,\hbar\omega\}$, respectively,
for a nucleus.
Integrating then over 
$\boldsymbol\alpha$ in Eq.~(\ref{dendef1}), 
one obtains using the
standard saddle-point method
\be\l{rhoE1Fnp}
\rho(E,{\bf Q}) \approx 
\frac{1}{(2\pi i)^{\kappa/2+1}}
\int d \beta~\beta^{\kappa/2}
\mathcal{J}^{-1/2}
\exp\left(\beta U + a/\beta\right)~,
\ee
where $U$ is the excitation energy,
\be\l{U}
U=E-E_0-\frac{1}{2}\Theta \omega^2~,
\ee
with $E_0=E^{(n)}_0+E^{(p)}_0$, and $\Theta=\Theta_{n}+\Theta_p$.
In Eqs.~(\ref{rhoE1Fnp}) and (\ref{denparnp}), 
\be\l{anp}
a=a_n+a_p, \qquad g=g_n+g_p~, 
\ee
and $a_\tau$ is
the $\tau$ component of the 
level density parameter, 
given approximately by Eqs.~(\ref{denparnp}) and (\ref{daFnp}).
The $\kappa$-dimensional 
 Jacobian determinant, $\mathcal{J}$, is defined 
 at the saddle point, 
 $\boldsymbol\lambda=\boldsymbol\lambda^\ast=\boldsymbol\alpha^\ast/\beta$,
Eq.~(\ref{Seqsdnp}), 
at a given $\beta$,
\be\l{Jacnp}
\mathcal{J}=\mathcal{J}\left(
      \frac{\partial \Omega}{\partial \boldsymbol\lambda},
      \boldsymbol\lambda\right)^\ast~,
\ee
where the asterisk indicates
the saddle point 
for the integration over $\boldsymbol\lambda$ at any $\beta$. 
    In the following, for simplicity of
    notations, we will omit the asterisk at $\boldsymbol\lambda^\ast$.
    For $\boldsymbol\lambda=\{\lambda_n,\lambda_p,\hbar\omega\}$ in
    the case of one of
    standard nuclear physics
    problems, ${\bf Q}=\{N,Z,M\}$,
    the Jacobian $\mathcal{J}$ [Eq.~(\ref{Jacnp})]
    is a simple three-dimensional
    determinant ($\kappa=3$). In the nuclear adiabatic approximation,
    where we may neglect
    the $\beta $ dependence of the level density parameter
    $a$ and of the
    moment of inertia $\Theta$
    in Eq.~(\ref{OmadFnp}),
    one has a diagonal form of the Jacobian:
    \be\l{Jacnpad}
\mathcal{J}\cong \hbar^{-2}\Theta~\mathcal{J}^{(2)}\left(
\frac{\partial \Omega}{\partial \lambda_n},
\frac{\partial \Omega}{\partial \lambda_p};
      \lambda_n,\lambda_p\right)~,
    \ee
    where the superscript in $\mathcal{J}^{(2)}$
    means the dimension 2
    of the determinant.
Within this approximation, taking 
the derivatives of Eq.~(\ref{OmadFnp}) for the potential $\Omega$
with respect to $\boldsymbol\lambda$
in the Jacobian $\mathcal{J}$, Eq.~(\ref{Jacnpad}),
up to linear terms in expansion over $1/\beta^2$ (Ref.~\cite{PRC}), one
obtains (see Ref.~\cite{IJMPE})\footnote{We shall present the Jacobian
  calculations for the main case
  of $\delta g<0$
  near the minimum of the level density and energy
  shell corrections, as mainly applied below.
  For the case of a
  positive $\delta g$ we change, for convenience, signs so that we will get
  $\xi>0$.}
\be\l{Jacnp2}
\mathcal{J}^{(2)}=
\mathcal{J}_0 \left(1+\xi\right),
\ee
where
\be\l{xiparnp}
\xi=\overline{\xi}/\beta^2,\qquad
\overline{\xi}=\mathcal{J}_2/\mathcal{J}_0~,
\ee
 with 
\be\l{J12np}
\mathcal{J}_0=
g_n(\lambda_n)g_p(\lambda_p),\quad
\mathcal{J}_2=
a^{\prime\prime}_pg_n(\lambda_n)+a^{\prime\prime}_ng_p(\lambda_p)~,
\ee
see Eqs.~(\ref{Jacnpad}), and (\ref{OmadFnp}).
 Up to a small asymmetry parameter squared, $~~X^2=$ $(N-Z)^2/A^2$, one has
approximately, 
$\lambda_n=\lambda_p=\lambda$,
and $~g_ng_p=g^2/4~$ [see Eq.~(\ref{anp}) for $g$]. 
Then, correspondingly, one can simplify
Eqs.~(\ref{J12np}) with (\ref{denparnp}) 
to have
\be\l{J02snp}
\mathcal{J}_0=
\frac14 g^2,\quad
\mathcal{J}_2=
\frac{\pi^2}{12}g^{\prime\prime}g~.
\ee

According to Eq.~(\ref{gdecomp}), a
decomposition of the Jacobian, Eq.~(\ref{Jacnp2}),
in terms of its 
smooth extended Thomas-Fermi 
and linear oscillating periodic-orbit 
components 
    of $\mathcal{J}_0$ and $\mathcal{J}_2$
can be found straightforwardly
with the help of 
Eqs.~(\ref{J12np}) and  (\ref{denparnp})
(see also Ref.~\cite{PRC}),
\be\l{Jdecomp}
\mathcal{J}_0=\tilde{\mathcal{J}}_0+\delta \mathcal{J}_0, \qquad 
\mathcal{J}_2=\tilde{\mathcal{J}}_2+\delta \mathcal{J}_2\approx \delta \mathcal{J}_2~.
\ee
 As demonstrated in Appendix \ref{appA}, the dominance
 of derivatives of the semiclassical expression 
 (\ref{goscsem})
 for the 
level density
shell corrections,
$\delta g_{\rm scl}$, in Eq.~(\ref{J02snp}) for
$\delta \mathcal{J}_2$, 
led to the
last approximation in Eq.~(\ref{Jdecomp}).
For smooth,
$\tilde{\mathcal{J}}^{(2)}$, and oscillating,
$\delta \mathcal{J}^{(2)}$,  components of
$\mathcal{J}^{(2)}\cong
\tilde{\mathcal{J}}^{(2)}+\delta \mathcal{J}^{(2)}$, Eq.~(\ref{J12np}),
one
finds with
the help of 
Eq.~(\ref{daFnp}),
\be\l{tilJnp}
\tilde{\mathcal{J}}^{(2)}\approx \tilde{\mathcal{J}}_0=
\tilde{g}_{n}\tilde{g}_{p}~,\quad
\delta \mathcal{J}^{(2)}= \delta \mathcal{J}_0 +\delta \mathcal{J}_2/\beta^2,
\ee
where $\tilde{g}_{\tau}\approx
g^{(\tau)}_{\rm{\tt{ETF}}}(\lambda_\tau)$ is approximately the 
(extended) Thomas-Fermi $\tau$ level-density component.
For linearized oscillating  major-shells components of
$\delta \mathcal{J}^{(\lambda)}$, 
one approximately arrives at
\be\l{dJnp}
\delta \mathcal{J}_0\approx 
\tilde{g}_{n}\delta g_p
    +\tilde{g}_{p}\delta g_n~,\quad
\delta \mathcal{J}_2\approx 
 - \frac{2\pi^4}{3}\left[
    \frac{\tilde{g}_{n} \delta g_p}{
      \mathcal{D}_{p}^2}
    +\frac{\tilde{g}_{p} \delta g_n}{
      \mathcal{D}_{n}^{2}}
    \right]~,
  \ee
  where $\delta g_\tau$ is the 
  periodic-orbit shell component,
  $\delta g_\tau\approx \delta g^{(\tau)}_{\rm scl}$
  [see Eq.~(\ref{goscsem})],
   $\mathcal{D}_{\tau} = \mathcal{D}^{(\tau)}_{\rm sh}=\lambda_\tau/A^{1/3}$ is
  approximately the distance between major (neutron or proton) shells
  given by
  Eq.~(\ref{periode}). 
   Again, up to  terms of the order of $X^2$,
   one simply finds from Eqs.~(\ref{tilJnp}), and (\ref{dJnp}),
    \be\l{J02dcompsnp}
    \tilde{\mathcal{J}}_0\approx \frac14\tilde{g}^2,\quad
   \delta \mathcal{J}_0\approx 
   \frac12\tilde{g}\delta g~,\quad
   \delta \mathcal{J}_2\approx
     -\frac{\pi^4}{3}\frac{g\delta g}{\mathcal{D}^2}~,
     \ee
where $\mathcal{D}=\lambda/A^{1/3}$.
Note that for
thermal excitations smaller or of the order of those of neutron resonances,
 the main contributions of the oscillating potential, 
$\delta \Omega_\tau$, and Jacobian, $\delta \mathcal{J}^{(2)}$, components
as functions of $\lambda_\tau$,
 are coming from the
differentiation of the sine function in
the PO 
level density  
component, $g^{(\tau)}_{\rm PO}(\lambda_\tau)$, Eq.~(\ref{goscsem}),
 through the PO
 action phase
$\mathcal{S}^{(\tau)}_{\rm PO}(\lambda_\tau)/\hbar$.
The reason is that, 
for large 
particle numbers, $A$, the
semiclassical large parameter,
$\sim\mathcal{S}^{(\tau)}_{\rm PO}/\hbar \sim A^{1/3}$, 
    leads to 
a dominating contribution, much larger than that coming
from  differentiation of other terms, such as,
the $\beta$-dependent function
$x^{(\tau)}_{\rm PO}(\beta)$, and 
the periodic-orbit 
period $t^{(\tau)}_{\rm PO}(\lambda_\tau)$.
Thus,  in the linear approximation over $1/\beta^2$,
we simply arrive 
to Eq.~(\ref{dJnp}), similarly to the derivations in 
    Refs.~\cite{PRC,IJMPE}.

    In the linear approximation in $1/\beta^2$, one finds
    from Eq.~(\ref{xiparnp}) for $\xi$ and
Eq.~(\ref{x}) for $x^{(\tau)}_{\rm PO}$, see also Eqs.~(\ref{tilJnp}), 
(\ref{dJnp}), and (\ref{anp}),
\be\l{xib}
\overline{\xi}
\approx \delta \mathcal{J}_2/\tilde{\mathcal{J}_0}\approx
\overline{\xi}_n
+\overline{\xi}_p~,\qquad \overline{\xi}_\tau=
-\frac{2\pi^4 \delta g_{\tau}}{3\tilde{g}_\tau\mathcal{D}^2_\tau}~,
  \ee 
 see also 
 Eq.~(\ref{periode}) for $\mathcal{D}^{}_\tau$.
 For convenience, 
 introducing 
 the dimensionless
energy shell correction,
$\mathcal{E}^{(\tau)}_{\rm sh}$, 
  in units of   
  the smooth extended Thomas-Fermi 
  energy per particle, $E^{(\tau)}_{\rm \tt{ETF}}/A$,
      one can present
      Eq.~(\ref{xib}) (e.g., for $\delta E<0$) as:
\be\l{xibdEnp}
\overline{\xi}_\tau
\approx \frac{4 \pi^6A^{1/3} \mathcal{E}^{(\tau)}_{\rm sh}}{3\lambda^2_\tau}~,
\quad 
\mathcal{E}^{(\tau)}_{\rm sh}=-\frac{A\delta E_\tau}{E^{(\tau)}_{\rm \tt{ETF}}}~.
\ee
The smooth
extended Thomas-Fermi 
energy $E^{(\tau)}_{\rm \tt{ETF}}$
       can be 
       approximated in order of magnitude, 
       as 
    $E^{(\tau)}_{\rm \tt{ETF}}\approx
    g^{(\tau)}_{\rm \tt{ETF}}(\lambda_\tau)\lambda_\tau^2/2 $~. 
    For a major shell structure, the energy
    shell correction, $\delta E=\delta E_n+\delta E_p$,
    is  expressed with
 a semiclassical accuracy, through the PO
 sum \cite{SM76,SM77,BB03,MY11,PRC,IJMPE}
      in Eq.~(\ref{dEPO0Fnp}) by
  $\delta E_\tau\approx \delta E^{(\tau)}_{\rm scl}$, where 
  (see Appendix \ref{appA}) 
  \be\l{dedgnp}
  \delta E^{(\tau)}_{\rm scl} 
  \approx
  \left(\frac{\mathcal{D}_{\tau}}{2 \pi}\right)^2~ \delta g_{\tau}(\lambda_\tau)~.
  \ee

    The correction, $\propto 1/\beta^4$, of
    the expansion in  $\propto 1/\beta^2$ of both 
        the potential shell correction, Eq.~(\ref{potoscparFnp})
   with Eqs.~(\ref{FESCFnp})-(\ref{x}), and the Jacobian,
    Eq.~(\ref{Jacnpad}), through the
    oscillating part, $\delta \mathcal{J}^{(2)}$,
    [see Eqs.~(\ref{tilJnp}) and (\ref{J02dcompsnp})]
 is relatively small
 for $\beta $ which, evaluated at the critical saddle-point
 values $T=1/\beta^\ast$, is related to the
  chemical potential $\lambda_\tau$
  as $T \ll \lambda_\tau$. Thus, the temperatures $T=1/\beta^\ast$,
when the saddle point $\beta=\beta^\ast$ exists, 
 are assumed to be 
much smaller than the chemical
potentials $\lambda_\tau$. The high order,
  $\propto 1/\beta^4$, term of this expansion  can be
  neglected under the following condition (subscripts $\tau$
  are omitted for a small asymmetry parameter $(N-Z)^2/A^2$,
  see also
  Ref.~\cite{PRC}): 
 \be\l{condUnp}
\frac{1}{\tilde{g}}\siml U\ll
\sqrt{\frac{90}{7}}\frac{a\lambda^2}{2\pi^4 A^{2/3}}~.
\ee
Using typical values for
parameters $\lambda\approx 40$ MeV
(1\,MeV\,=\,1.60218$\times 10^{-13}$\,joules), $A=200$, and
$a =A/K\sim 20 $~MeV$^{-1}$ ($K\sim 10$~MeV), $1/\tilde{g}\sim 0.1-0.2$~MeV,
 one finds, 
numerically, that
the r.h.s. of this inequality is of the order of the chemical potential, $\lambda$, 
see Ref.~\cite{KS18}. Therefore, one obtains approximately
$U \ll \lambda$.
For simplicity,
          the small shell and temperature corrections to
        $\lambda_\tau(\mathcal{N}_\tau)$ obtained  from the conservation
          equations,  Eq.~(\ref{Seqsdnp}),
          can be neglected.
           Using the linear shell correction approximation 
           of the leading order \cite{BD72}
           and constant particle-number density of symmetric nuclear matter, 
           $\rho^{}_0=2k_F^{3}/3\pi^2=0.16$~fm$^{-3}$
 (1\,fm\,=\,$10^{-15}$\,meters)
           ($k^{{n}}_F\approx k^{{p}}_F\approx k_F=1.37$ fm$^{-1}$
           is the Fermi momentum in units of $\hbar$),
           one finds about a constant value for 
           the chemical potential,
           $\lambda_\tau \approx \hbar^2k^{2}_F/2m \approx 40$ MeV,
           where $m$ is the nucleon mass.
       In the derivations of the condition (\ref{condUnp}),
       we used the POT 
       distances between major shells,
       $\mathcal{D}^{(\tau)}_{\rm sh}$, 
       Eq.~(\ref{periode}),
       $\mathcal{D}^{(\tau)}_{\rm sh}\approx \lambda/A^{1/3}$.
Evaluation of the upper limit for the
excitation energy at the saddle point $\beta=\beta^\ast=1/T$
is justified because: this upper
limit is always so large that this point does  certainly exist.
Therefore, for 
 consistency, one can neglect the
  quadratic, $1/\beta^2$ (temperature $T^2$), corrections to the Fermi
  energies $\varepsilon^{(\tau)}_{F}$ in the chemical
  potential\footnote{In
    our semiclassical picture,
          it is convenient to determine  
          the Fermi energy,  $\varepsilon^{(\tau)}_{F}$,
          to be counted from 
          the bottoms 
          of the neutrons and protons potential wells.}
  $\lambda_\tau\approx \varepsilon^{(\tau)}_{F}$ (or,  
     $\lambda_\tau \approx \lambda \approx \varepsilon^{}_{F} $), for large 
  particle numbers $A$ and small asymmetry parameter $X^2$.

  \subsection{Shell and isotopic asymmetry effects within the standard
    saddle-point method}\label{subsec2-2}
  
For simplicity, one can start with a  
direct application of the standard 
saddle-point approach
for calculations of the inverse Laplace integral over $\beta$ in Eq.~(\ref{rhoE1Fnp}).
In this 
way (Appendix \ref{appB}), including
    the nuclear shell (Ref.~\cite{PRC}) and isotopic-asymmetry
    (Ref.~\cite{IJMPE}) effects, one arrives at 
\be\l{SPMgennp}
\rho(E,N,Z,M)\approx
\frac{\sqrt{2}~\hbar~a~\exp(2\sqrt{aU})}{8\pi^3U^{3/2}\sqrt{\mathcal{J}_0(1+\xi^\ast)\Theta}}
\approx
\frac{\sqrt{2}~\hbar~\exp(2\sqrt{aU})}{24\pi U^{3/2}\sqrt{(1+\xi^\ast)\Theta}}~.
\ee
 Here, $a=\pi^2 g(\lambda)/6$ is the total level density parameter,
    Eqs.~(\ref{anp}) 
and (\ref{daFnp}), $\mathcal{J}_0$ is the component of the
Jacobian $\mathcal{J}^{(2)}$,
Eq.~(\ref{Jacnp2}), which is independent of
 $\beta$ but depends on the shell structure, 
  and
\be\l{pars}
\xi^\ast=\frac{\overline{\xi}}{\beta^{\ast\;2}}
\approx\frac{2\pi^4 U }{3a A^{1/3}}\mathcal{E}_{\rm sh}^{\prime\prime}(\lambda)
  \approx\frac{8\pi^6 U A^{1/3}}{3a\lambda^2}\mathcal{E}_{\rm sh}~.
\ee
 The relative shell correction,
    $\mathcal{E}_{\rm sh} \approx \mathcal{E}_{\rm sh}^{(\tau)}$, given
    by Eq.~(\ref{xibdEnp}), 
    is almost independent of $\tau$ for small $X^2=[(N-Z)/A]^2$,
see also Eqs.~(\ref{xib}) and (\ref{dedgnp}). 
 The asterisk means $\beta=\beta^\ast=\sqrt{a/U}$ at the saddle point.
 In the second equation of (\ref{SPMgennp}), and of (\ref{pars})
 we used
    $\lambda_n\approx \lambda_p\approx \lambda$
 for a small asymmetry parameter, $X^2$, 
 together with Eq.~(\ref{d2Edl2}) for the
derivatives of the energy shell corrections 
    and Eq.~(\ref{periode}) for
the mean distance between
neighboring major shells near the Fermi surface,
$\mathcal{D}\approx \lambda/A^{1/3}$.

In Eq.~(\ref{SPMgennp}), the quantity $\xi^\ast$ 
is $\xi$ in
    Eq.~(\ref{xiparnp}), 
taken at the saddle point,
    $\beta=\beta^\ast$ ($\lambda_\tau=\lambda_\tau^\ast$). 
 This quantity
is the sum of the two $\tau$ contributions, $\xi^\ast=\xi^\ast_n
+\xi^\ast_p$, $~~\xi^\ast_n\approx \xi^\ast_p$.  The value of $\xi^\ast$ 
is
 approximately proportional to the excitation energy, $U$, and
to the  relative 
energy shell corrections, $\mathcal{E}_{\rm sh}$, 
Eq.~(\ref{xibdEnp}), and inversely proportional to the level density
parameter, $~a~$,
 with 
$~\xi^\ast\propto U A^{1/3}\mathcal{E}_{\rm sh}/(a \lambda^2)$. 
 For typical parameters
  $\lambda=40$ MeV, $A\sim 200$,
  and 
  $\mathcal{E}_{\rm sh}=|\delta E~A/E_{\rm \tt{ETF}}|\approx 2.0$
  \cite{BD72,MSIS12},
  one finds the estimates
  $\xi^\ast\sim 0.1 - 10$ for temperatures $T \sim 0.1-1$ MeV. 
      This corresponds approximately
  to rather a wide 
  excitation energy region, $U=0.2-20$~MeV for the inverse level
  density parameter, $K$, $K=A/a=10$~MeV; see
  Ref.~\cite{KS18}
  ($U=0.1-10$~MeV for $K=20$~MeV). This energy range includes the 
     low-energy states and states
   significantly
  above the neutron resonances.
  Within the 
  periodic-orbit theory \cite{SM76,BB03,MY11} 
  and 
  extended Thomas-Fermi approach \cite{BG85,BB03,KS20},
    these values are given finally by using the realistic
    smooth energy $E_{\rm \tt{ETF}}$ for which
    the binding energy \cite{MSIS12}   
    is approximately
    $E_{\rm \tt{ETF}}+ \delta E$.

    Eq.~(\ref{SPMgennp})
     is a more general shell-structure Fermi-gas (SFG) 
    asymptote,
    at large excitation energy,
  with respect to the well-known \cite{Be36,Er60,GC65}
   Fermi gas (FG) 
  approximation for $\rho(E,N,Z,M)$, which is equal to that of
  Eq.~(\ref{SPMgennp}) at
  $\xi^\ast \rightarrow 0$,
  \be\l{FG}
  \rho(E,N,Z,M)\rightarrow
  \frac{\sqrt{2}~\hbar~\exp(2\sqrt{aU})}{24\pi U^{3/2}\sqrt{\Theta}}~.
\ee
Notice that a shift of the inverse level-density parameter
$K$ due to shell effects with increasing excitation energies
which is related to temperatures of the order of 1-3 MeV is
discussed in Refs.~\cite{PRC,SN90,SN91}.

\begin{figure}
    \includegraphics*[width=8cm,clip]{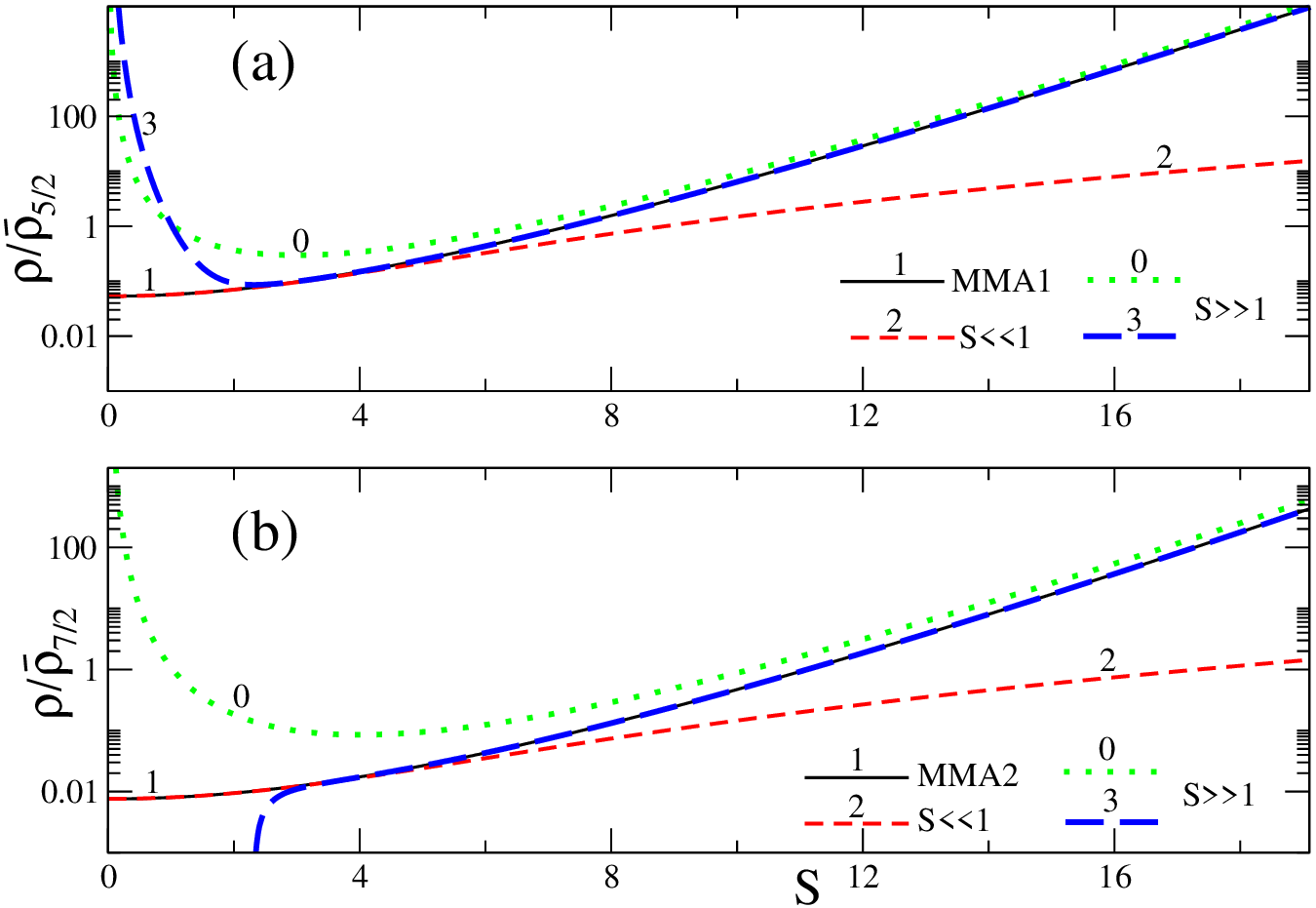}  
\caption{
    Level density $\rho$ [Eq.~(\ref{denbesnp})], 
    in units of $\overline{\rho}_\nu$, with the accurate result ``1''
    (solid line),
    Eq.~(\ref{rho1}), $(a)$
    for $\nu=5/2$ [MMA1 (i)], and $(b)$, Eq.~(\ref{rho2}),
    for $\nu=7/2$ [MMA2 (ii)], 
    shown as
    functions of the entropy $S$ for different approximations:
     (1) $S \ll 1$
    (dashed lines),  Eq.~(\ref{den0gennp})
    at the second order, and (2) $S \gg 1$
    (dotted and 
    rare-dashed lines),~ 
       Eq.~(\ref{rhoasgennp}); 
       ``0'' is the main term of the expansion in powers of $1/S$,
     and ``3''
     is the expansion  over $1/S$ 
     up to 
     first [in $(a)$], and second [in $(b)$]
    order terms in square brackets of Eq.\ (\ref{rhoasgennp}),
    respectively.}
\label{fig1}
\end{figure}

\subsection{Shell, isotopic-asymmetry and rotational effects within the MMA}
\label{subsec2-3}

Under 
  the condition of Eq.~(\ref{condUnp}),
one can obtain simple analytical expressions for the level density
$\rho(E,{\bf Q})$, beyond the standard saddle-point method
of both the previous subsection and Appendix \ref{appB},
Eq.~(\ref{SPMgennp}).
Using the integral representation
(\ref{rhoE1Fnp}) in the adiabatic approximation, one can simplify significantly
 the square root
 Jacobian factor $\mathcal{J}^{-1/2}$ [Eq.~(\ref{Jacnpad})]  
by its expansion\footnote{At each finite order of these expansions, one can  
accurately take \cite{PRC}
the inverse Laplace transformation.
Convergence of the corresponding corrections to the level density,
Eq.~(\ref{rhoE1Fnp}),
after applying this inverse
transformation,
can
be similarly proved as carried out in Ref.~\cite{PRC}.} 
over small values of $\xi$ or of $1/\xi$;
see Eq.~(\ref{Jacnp2}).
Expanding now this Jacobian factor
    at linear order in $\xi$ and $1/\xi$, 
one arrives at two different approximations marked below by cases
(i) and (ii), respectively. 
 Then, taking the
inverse Laplace transformation over 
$\beta$   in Eq.~(\ref{rhoE1Fnp}),
more accurately
(beyond that by the standard saddle-point method), 
one approximately obtains (see Refs.~\cite{PRC,IJMPE}),
    \bea\l{denbesnp}
    &\rho \approx \rho^{}_{\tt{MMA}}(S)
    =\overline{\rho}_\nu~f_\nu(S)~,\\
    &~~~~~~f_\nu(S)=
  S^{-\nu}I_{\nu}(S)~,\quad S=2\sqrt{a U}~. \l{f}
  \eea
  %
    Here, $I_\nu(S)$ is the modified Bessel function with the index $\nu=\kappa/2+1$
    for case (i) and $\nu=\kappa/2+2$ for case (ii),
    which is determined by the number of the additional integrals of motion above the energy
    $E$, i.e., the dimension $\kappa$ of the vector ${\bf Q}$.
    The expression for the entropy $S$ in the argument of this function, $I_\nu(S)$,
    is given by Eq.~(\ref{entnp}) at the saddle point $\boldsymbol\lambda=\boldsymbol\lambda^\ast$.
    This expression happens to be  
    similar to that of the ideal Fermi gas model,
    but the level density parameter $a$ in this entropy $S$
    is a sum of two semiclassical terms. One of them is the
    extended Thomas-Fermi term, beyond
    the Fermi gas approach. Another term is
    the periodic-orbit shell correction of the Strutinsky's method \cite{BD72};
    see, e.g., Eqs.~(\ref{anp}) and (\ref{denparnp}) in the adiabatic
    approximation. The
    excitation energy $U$ is given by Eq.~(\ref{U}) with the explicit
    dependence on 
   the shell correction $\delta E$ and the rotational frequency $\omega$.  
    For small
        $\xi \sim \xi^\ast \ll 1$, we have case (i), and for
        large $\xi \sim \xi^\ast \gg 1$, we have
        case (ii), where $\xi^\ast$ is, thus,
    the critical shell-structure quantity,
    given by expression of Eq.~(\ref{pars}).

    For the level density $\rho(E,N,Z,M)$ of nuclear spectra,
             one finds
    $\nu=5/2$  for case (i) and $7/2$ for case (ii), 
             respectively. In these cases, the modified Bessel
             function $I_\nu(S)$ are expressed
             in terms of the elementary functions, $\sinh(S)$ and $\cosh(S)$.           
     In cases (i) and (ii),
    named below as the MMA1 and MMA2 approaches,  respectively,
    one obtains
    Eq.~(\ref{denbesnp})
    with different coefficients $\overline{\rho}_\nu$ 
    (see also Refs.~\cite{PRC,IJMPE}),
\bea\l{rho1}
&\rho^{}_{1}(E,N,Z,M)\cong \rho^{}_{\tt{MMA1}}(S)=
\overline{\rho}_{5/2}S^{-5/2}I_{5/2}(S),\\
&\overline{\rho}_{5/2}=\frac{(2 a)^{5/2}\hbar}{\sqrt{\mathcal{J}_0 \Theta}}\approx
\frac{4\pi^2\hbar\sqrt{2} a^{3/2}}{3 \Theta^{1/2}}~,
\qquad \mbox{MMA1 (i)},\l{rhobar1}\\
&\rho^{}_2(E,N,Z,M)\cong\rho^{}_{\tt{MMA2}}(S)=\overline{\rho}_{7/2}S^{-7/2}I_{7/2}(S),\l{rho2}\\
&\overline{\rho}_{7/2}\approx 
\frac{(2a)^{7/2}\hbar}{ \sqrt{\overline{\xi}\mathcal{J}_0 \Theta}}
\approx \frac{8\pi^2 \sqrt{2}a^{5/2}}{3\sqrt{\overline{\xi}\Theta}}~, \qquad \mbox{MMA2 (ii)},
\l{rhobar2}~
\eea
where $\overline{\xi}$ and $\mathcal{J}_0$ are given by
Eqs.~(\ref{xiparnp}) and
(\ref{J12np}), respectively; see also Eq.~(\ref{J02snp})
    for small asymmetry parameter, $X^2 \ll 1$.
Eqs.~(\ref{rho1}) and (\ref{rho2}) depend explicitly on the moment of
inertia $\Theta$, which is the sum of the ETF
component $\Theta_{\tt{ETF}}$ and
shell corrections $\delta \Theta$, $\Theta=\Theta_{\tt{ETF}} +\delta \Theta$. 
For the 
Thomas-Fermi approximation to the coefficient $\overline{\rho}_{7/2}$
within the case (ii) 
one finds \cite{NPA,PRC}
\bea\l{rho2b}
&\rho^{}_{2b}(E,N,Z,M)\cong \rho^{}_{\tt{MMA2b}}(S)=\overline{\rho}^{(2b)}_{7/2}S^{-7/2}I_{7/2}(S),\\
&\overline{\rho}^{(2b)}_{7/2}\approx \frac{8\pi\sqrt{3}\lambda a^{5/2}\hbar}{3 \sqrt{\Theta}}~, \qquad \mbox{MMA2b (ii)}.
\l{rhobar2b}
\eea
In the derivation of the coefficient, $\overline{\rho}^{(2b)}_{7/2}$,
we assume in
    Eq.~(\ref{rhobar2b}) for $\overline{\rho}_{7/2}$
that the magnitude
of the relative shell corrections
    $\mathcal{E}_{\rm sh}$, $\overline{\xi}\propto \mathcal{E}_{\rm sh}$
    [see Eqs.~(\ref{xiparnp}), 
    (\ref{J02snp}), and (\ref{xibdEnp})]
are  extremely
    small but their derivatives yield large
    contributions through the 
    level density derivatives  $g^{\prime\prime}(\lambda)$,
$g \propto A/\lambda$, 
  as in the 
 Thomas-Fermi approach.
For large entropy
$S$, one finds from Eq.~(\ref{f})
 \begin{equation} 
 f_\nu(S) =\frac{\exp(S)}{S^{\nu}\sqrt{2\pi S}}\left[1+\frac{1-4\nu^2}{8S}
    +\mbox{O}\left(\frac{1}{S^2}\right)\right].
 \label{rhoasgennp}
\end{equation}
 The same leading results in the expansion (\ref{rhoasgennp}) 
for $\nu=5/2$ (i) and  $\nu=7/2$ (ii) 
at large excitation energies $U$
are also derived 
from the shell-structure Fermi gas  
formula (\ref{SPMgennp}).  
     At small entropy, $S \ll 1$, one also obtains
     from Eq.~(\ref{f})
     the finite combinatorics power
expansion \cite{St58,Er60,Ig72}: 
\begin{equation}
    f_\nu(S)=
  \frac{2^{-\nu}}{\Gamma(\nu+1)}\left[1+\frac{S^2}{4(\nu+1)}+
    \mbox{O}\left(S^4\right)\right],
\label{den0gennp}
\end{equation}
where $\Gamma(x)$ is the gamma function.
    This expansion over powers of 
    $S^2 \propto U$ 
    is the same 
    as that of  the 
    constant ``temperature'' model \cite{GC65,ZK18,ZH19,KZ20},    
    used often for the level density calculations,
    but here, as in Refs.~\cite{PRC,IJMPE}, we have it without
free fitting parameters.
%
\begin{figure}
  \centerline{\includegraphics*[width=8cm,clip]{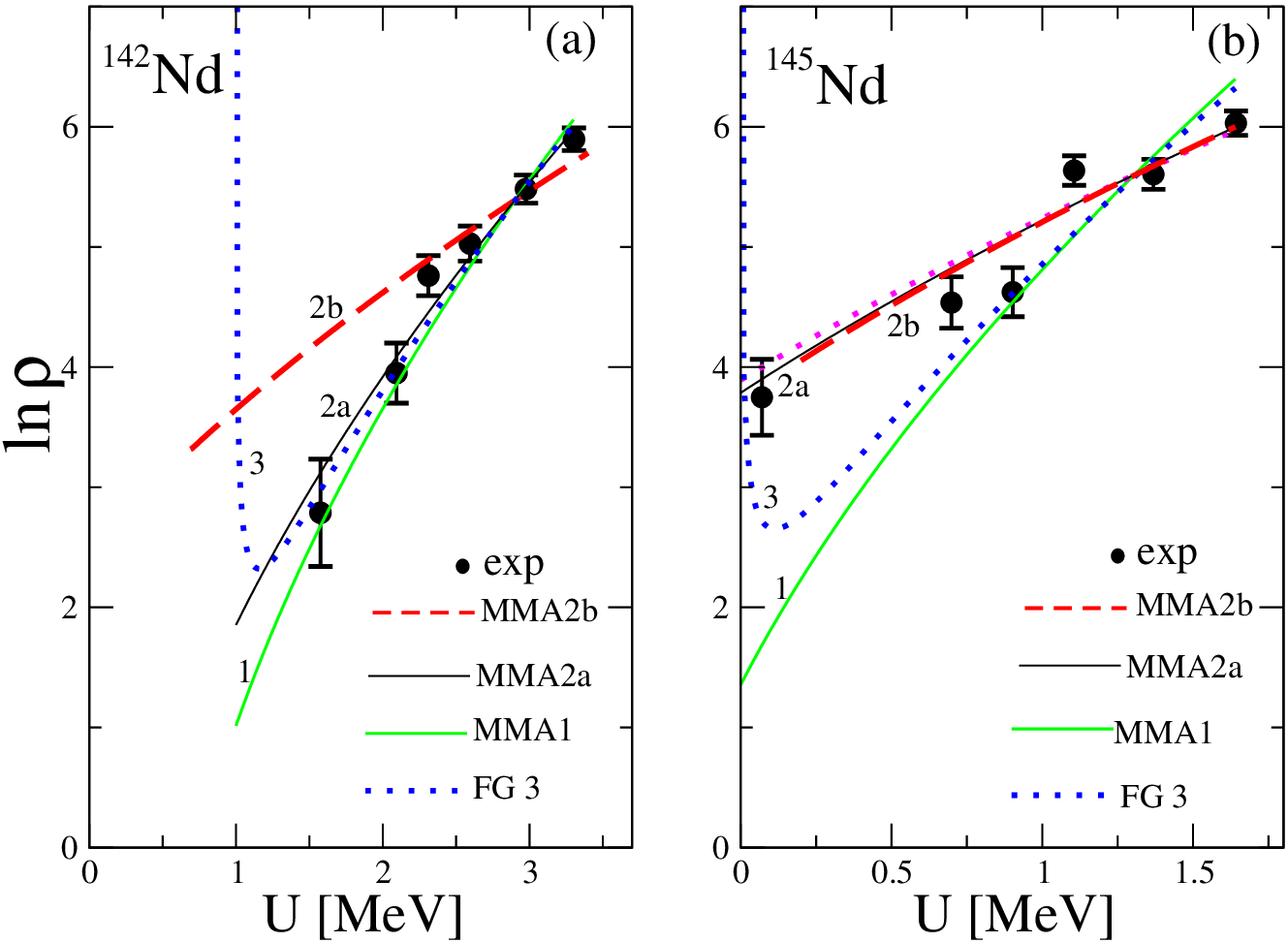}}
 \caption{ 
   Level density, $\mbox{ln}\rho(E,N,Z)$,
    for
   low energy states in nuclei $^{142}$Nd (a),
   and $^{145}$Nd (b)
   were calculated  within different
    approximations: The MMA 
    solid green ``1'' and black ``2a'' lines,
    Eqs.~
    (\ref{rho1tot})
    and (\ref{rho2tot}),
    at the  realistic relative shell correction \cite{MSIS12}
    $\mathcal{E}_{\rm sh}$, respectively; the MMA dashed red line
    ``2b'',
    Eq.~(\ref{rho2btot}),
        and
        the 
        Fermi-Gas  
 rare blue dotted line,
      Eq.~(\ref{FGtot}),
      approaches are presented.
      The realistic values of $
     \mathcal{E}_{\rm sh}$= 0.44 (a), and 0.25 (b) for MMA2 are    
     taken from Ref.~\cite{MSIS12}
      (with chemical potentials $\lambda_n\approx\lambda_p\approx
     \lambda$, where
      $\lambda=40$ MeV).
     Experimental dots with error bars
     are obtained 
     from the ENSDF database \cite{ENSDFdatabase} 
    by using the sample method \cite{So90,IJMPE}. 
}
\label{fig2}
\end{figure}

 In contrast to the finite MMA limit (\ref{den0gennp}) for the level density,
Eq.~(\ref{denbesnp}), the asymptotic SFG [Eq.~(\ref{SPMgennp})]
and FG [Eq.~(\ref{FG})] expressions are obviously divergent
at $U\rightarrow 0$.
 Notice also that the MMA1 approximation for the level density,
  $\rho(E,N,Z,M)$,
  Eq.~(\ref{rho1}),
  can be applied for
  large excitation energies, $U$, with respect to the
  collective rotational excitations. This also the case for the
  shell-structure Fermi gas (SFG) and Fermi gas (FG) approximations 
  if one can neglect shell effects, $\xi^{\ast}\ll 1$.
  Thus, the level density $\rho(E,N,Z,M)$ in the case (i), Eq.~(\ref{rho1}),
  has a wider range of applicability over the excitation energy variable
  $U $ than the MMA2 case (ii) [Eq.~(\ref{rho2})].
  The MMA2 approach has, however, another advantage 
  of describing the important shell structure effects. 
      The main effects of
  the interparticle
  interaction, statistically averaged over particle numbers,
  beyond the shell correction of the mean field, within the Strutinsky's
  shell correction 
      method, was taken into account by the extended Thomas-Fermi
  components of MMA expression
  (\ref{denbesnp}) for the level density, $\rho(E,N,Z,M)$.
  These components of $\Omega$ and $a$,
  Eqs.~(\ref{OmFnp}) and (\ref{denparnp}), respectively,
  are given by
  the extended Thomas-Fermi potential,
  $\Omega^{}_{\rm \tt{ETF}}$, Eq.~(\ref{TFpotF}),
  and the level-density parameter, $a^{}_{\rm \tt{ETF}}$,
  Eq.~(\ref{daFnp}).

 In Fig.~\ref{fig1} we show
the level density dependence
$\rho(S)$, Eq.~(\ref{denbesnp}), 
for $\nu=5/2$ in $(a)$ and $\nu=7/2$ in $(b)$, 
on the
entropy variable $S$ with
the corresponding asymptote. In this figure, 
a small [$S\ll 1$, Eq.~(\ref{den0gennp})] and
large [$S\gg 1$, Eq.~(\ref{rhoasgennp})] entropy $S$ behavior 
is presented. For small $S\ll 1$ expansion we take into account the quadratic
approximation ``2'', where $S^2 \propto U$, that is  the same as in the linear
expansion within the
CTM \cite{GC65,ZK18,ZH19}. For large $S\gg 1 $
we neglected the corrections of the inverse power entropy expansion of the
pre-exponent factor
in square brackets of Eq.~(\ref{rhoasgennp}), lines ``0'', and
took into account the 
corrections of the
first [$\nu=5/2$, $(a)$] and up to second [$\nu=7/2$, $(b)$]
    order in $1/S$  (rare dashed lines ``3'') to show 
their slow convergence
to the accurate 
MMA result ``1'' (\ref{denbesnp}).
It is interesting to find almost a 
constant shift of the results of approximation at
large $S$ (dotted line ``0'') with the simplest,
$\rho \propto \exp(S)/S^{\nu+1/2}$,
asymptotic saddle point method (SPM) in respect
to the accurate  
MMA results of Eq.~(\ref{denbesnp}) (solid line ``1'').
This may clarify
one of the phenomenological
models, e.g., the 
back-shifted Fermi-gas 
(BSFG) model for the level density \cite{DS73,So90,EB09}.

  \section{Spin-dependences and total
    level densities}
  \l{sec3}

  Assuming that there are no external forces acting on a
   nuclear system,
  the total angular momentum, ${\bf I}$, and its projection $M$ on a
  space-fixed axis 
  are conserved, and states with a given energy, $E$,
  and spin, $I$, are $2I\!+\!1$ degenerate.
  In this case we use Eq.~(\ref{rho1}) for
  $\rho^{}_1(E,N,Z,M)$ in the case MMA1 (i) and
  Eq.~(\ref{rho2}) for
  $\rho^{}_2(E,N,Z,M)$ [or $\rho^{}_{2b}(E,N,Z,M)$, Eq.~(\ref{rho2b})]
  in the case MMA2 (ii) [MMA2b (ii)]. In the first part of this section we will
  consider the ``parallel''
  rotation,  i.e., an alignment of the
  individual  angular momenta of the particle
  along the axis $0z$ of arbitrary
  space-fixed coordinate system (see POT shell-structure
      level density in
      Refs.~\cite{KM79,PRC,MK79}). The second part is devoted to the 
      collective rotation around the axis $0z$, perpendicular
      to the symmetry axis $0z$
      \cite{Bj74,BM75,Ig83}.
          This section will be ended by the total integrated
      level-density MMA approach.
      
\subsection{``Parallel'' rotations}
\l{subsub3-1}

      For the ``parallel''
      rotation around the axis $0z$ of the space-fixed coordinate system
      (Refs.~\cite{KM79,PRC,MK79}),     
  %
      one can 
   use Eqs.~(\ref{rho1}), (\ref{rho2}), and (\ref{rho2b}) for the level density
   $\rho(E,N,Z,M)$, where $M$ is the projection of the angular momentum
   to the axis $0z$.
     The argument of the Bessel function, $f_{\nu}(S)\propto I_\nu(S)$,
      is the
  entropy $S(E,N,Z,M)=2\sqrt{aU}$,
  with  the $M$ dependent excitation energy $U$ [Eq.~(\ref{U})].
    Indeed, in the 
  adiabatic mean-field approximation, 
  the POT level density parameter $a$ 
is given by Eqs.~(\ref{anp}) and (\ref{daFnp}). 
For the intrinsic excitation energy $U$ [Eq.~(\ref{U})], 
one finds
\bea\l{Eex}
 &U=E-E_{0}-E_{\rm rot}~,\\
&  E_{\rm rot}=E^{\parallel}_{\rm rot}(M)=
\frac{\hbar^2M^2}{2\Theta},\quad M=\frac{\Theta}{\hbar} \omega~,
  \l{Erorpar}
  \eea
 where, $E_{0}=\tilde{E} +\delta E$,
  is the same intrinsic (non-rotating) shell-structure energy, as in
  Eq.~(\ref{OmadFnp}). 
 With the help of the conservation equations (\ref{Seqsdnp}) 
 for the  saddle point,
we deduced the value of the rotation frequency $\omega$,
using the second equation in Eq.~(\ref{Erorpar}). For the
 moment of inertia (MI)
  $\Theta$ with respect to the axis $0z$ in Eq.~(\ref{Eex}),
  one has a similar SCM decomposition:
  \be\l{MI}
  \Theta=\tilde{\Theta} + \delta \Theta~.
  \ee
  Here, $\tilde{\Theta}$ is the (E)TF MI
      component which can be approximated by the
     (E)TF expression, Eq.~(\ref{rigMIpar}), and $\delta \Theta$ is the MI
      shell correction for the axially symmetric mean field. 
     This MI can be presented explicitly analytically
      for the spherically symmetric mean field
      by Eq.~(\ref{dMIsph}).
   In
   Appendix \ref{appA} we present the specific POT derivations by assuming
  a spherical symmetry of the mean field potential, as full analytical example;
  see Eq.~(\ref{potoscpar}) with Eq.~(\ref{dMIsph})
  for the potential shell correction 
$\delta \Omega(\beta,\lambda,\omega)$ ($\lambda_n\approx \lambda_p\approx \lambda$).


 It is common to use in applications 
\cite{Be36,Er60,BM67} of the level density its
    dependence on the spin $I$,
$\rho(E,N,Z,I)$.
    In this subsection, we will consider 
    only the academic axially-symmetric potential
case which can be realized practically for
the spherical  or axial symmetry of a mean nuclear field.
Using 
Eqs.~(\ref{rho1}), (\ref{rho2}), and (\ref{rho2b}),
under the same assumption of a
closed rotating system 
and, therefore,  with
conservation of the integrals of motion,  the spin $I$
and its projection $M$ on the space-fixed axis,
one can calculate the corresponding
 spin-dependent level density $\rho(E,N,Z,I)$ 
  for a
  given energy $E$, neutron $N$ and proton $Z$ 
  numbers, 
  and total
angular momentum $I$ 
 by employing  the Bethe formula \cite{Be36,BM67,Ig83,So90},
\bea \l{denEAIgen}
\rho(E,I)&=&\rho(E,M=I) - \rho(E,M=I+1)\nonumber\\
&\approx&
-\left(\frac{\partial \rho(E,M)}{\partial M}\right)^{}_{M=I+1/2}~.
\eea
Here and in the following we omit for
simplicity the common arguments $N$ and $Z$.
For this
level density, $\rho(E,N,Z,I)$, 
one 
obtains approximately from 
Eqs.~(\ref{rho1}), (\ref{rho2}), (\ref{rho2b}), and (\ref{Eex}),
\be 
\rho(E,I) = \rho^{}_{\rm \tt{MMA}}(E,I) \approx
\frac{a\overline{\rho}_{\nu}\hbar^2(2I+1)}{\Theta}f_{\nu+1}(S)~.
\l{denbesI}
\ee
Here, $S$ is the same entropy [$S=2\sqrt{aU}$ from Eq.~(\ref{f})] given by
Eq.~(\ref{entnp}) at the saddle point
$\boldsymbol\lambda=\boldsymbol\lambda^\ast$, $a$ is the level density
parameter [Eqs.~(\ref{anp}) and
  (\ref{denparnp}); see also its $\omega^2$ correction
  in Eq.~(\ref{a0par}) for the spherical case], 
$U$ is the
excitation energy (\ref{Eex}), and $\nu$ equals 5/2 and 7/2,
in Eq.~(\ref{rho1}), and Eqs.~(\ref{rho2}), and (\ref{rho2b}), respectively.
The multiplier $2I+1$ in
Eq.~(\ref{denbesI}) appears because of the substitution $M=I+1/2$ into
the derivative in
Eq.~(\ref{denEAIgen}). In order to obtain
the approximate MMA total level density $\rho(E,N,Z)$
from the spin-dependent level density $\rho(E,N,Z,I)$, one
can multiply
Eq.~(\ref{denbesI}) by the spin
 degeneracy factor $2 I+1$ and integrate (sum)
 over all spins $I$,
 \be\l{totLD}
\rho(E)=\sum_{i}(2I_i+1)~\rho(E,I_i) \approx \int d I~(2I+1)~\rho(E,I)~.
 \ee

Using the expansion of the Bessel functions in Eq.~(\ref{denbesI})
over the argument $S$
for $S\ll 1$ [Eq.~(\ref{den0gennp})],
one finds a  combinatorics expression. 
For large $S$ [large excitation energy, $aU\gg 1$,
Eq.~(\ref{rhoasgennp})], one obtains from
Eq.~(\ref{denbesI}) the asymptotic Fermi gas expansion.
Again, the main term in the expansion for large $S$,
Eq.~(\ref{rhoasgennp}), coincides with the full SPM limit to the
inverse Laplace integrations in Eq.\ (\ref{dendef1}).
For small angular momentum $I$ and large excitation energy $U_0=E-E_{0}$, 
so that,
\be\l{Iexp}
\frac{E_{\rm rot}(I)}{U_0}\approx\frac{\hbar^2I(I+1)}{2\Theta~U_0} \ll 1~,
\ee
one finds the standard separation of the level density,
 $\rho^{}_{\rm \tt{MMA}}(E,N,Z,I)$, into the
product of the dimensionless
spin-dependent Gaussian multiplier, $\mathcal{R}(I)$, and 
another spin-independent factor. Finally, 
one finds 
\begin{equation}
  \rho^{}_{\rm \tt{MMA}}(E,I) \approx
 \frac{\overline{\rho}^{}_{\nu}~ \mathcal{R}(I)~\exp\left(S_0\right)}{
2~S_0^{\nu-1}\sqrt{2\pi S_0}}~,\quad S_0=2\sqrt{aU_0}~, 
\label{rhoIexp2}
\end{equation} 
where $\nu=5/2$ and $7/2$ for cases (i) and (ii), respectively.
The Gaussian spin-dependent factor $\mathcal{R}(I)$ is given by
\begin{equation}
\mathcal{R}(I)=\frac{2I+1}{q^2}
\exp\left(-\frac{I(I+1)}{2q^2}\right),~~ q^2=\frac{\Theta}{\hbar^2}\sqrt{\frac{U_0}{a}}~,
\label{qIexp}
\end{equation} 
where $q^2$ is the dimensionless spin
dispersion. This dispersion $q^2$ at the saddle point,
$\beta^\ast=1/T=\sqrt{a/U_0}$, is the standard spin dispersion
$\Theta T/\hbar^2$, see Refs.~\cite{Be36,Er60}.
%
Note that the power dependence of the pre-exponent factor of the
   level density $\rho(E,I) $ on  the
   excitation  energy, $U_0=E-E_0$, differs from that of $\rho(E,M)$;
   see Eqs.~(\ref{rho1}), (\ref{rho2}), and (\ref{rhoIexp2}). 
   The exponential
   dependence, $\rho \propto \exp[2\sqrt{a(E-E_0)}]$, for large excitation
   energy $E-E_0$ is the same for $\nu=5/2 $ (i) and $7/2$ (ii),
   also for any $\nu$ but the pre-exponent factor for the case of
   $\nu=5/2$ is different than that for the case of $7/2$, see
   Eq.~(\ref{rhoIexp2}).
    A small angular momentum $I$ means that the 
       condition of Eq.~(\ref{Iexp})  
       was applied.  Eq.\ (\ref{rhoIexp2}) 
       with
   Eq.~(\ref{qIexp}),  are valid 
   for excited states 
    within approximately the condition 
   $1/\tilde{g} \ll U \ll \lambda$,  see
    Eq.~(\ref{condUnp}).
     For relatively small
    spins [Eq.\ (\ref{Iexp})]  we have 
    the so-called 
    small-spins Fermi-gas
model (see, e.g., Refs.\ \cite{Be36,Er60,GC65,BM67,Ig83,So90,KS20}).

\subsection{``Perpendicular'' collective rotations}
\l{susub3-2}

Equations~(\ref{rho1}), (\ref{rho2}), and (\ref{rho2b}) for the
 level density $\rho(E,M)$
with the projection $M$ of the angular momentum ${\bf I}$
    can be used for the calculations of
    the level density 
$\rho(E,N,Z,\mathcal{K})$, where $\mathcal{K}$ is the specific projection of 
    ${\bf I}$ on the symmetry axis of the axially symmetric potential
    \cite{Ig83,Bj74,BM75,MK79}
    (K in notations of Ref.~\cite{BM75}, and $\mathcal{K}$ here
    should not be confused with the inverse level-density
    parameter).
General derivations of these equations applicable for axially symmetric systems
in
the previous part of this section are specified for a ``parallel'' rotation
and its basic characteristics are presented in Appendix \ref{appA} by
using the spherical potential.
However, the results for the
spin-dependent level
density, $\rho(E,I)$ 
[Eqs. (\ref{denbesI})-(\ref{rhoIexp2})] cannot be immediately
applied for comparison with the available experimental data on
rotational bands  in the collective rotation of a deformed nucleus.
They are 
studied within the unified rotation model \cite{BM75}
in terms of the spin $I$ and its projection $\mathcal{K}$
to the internal symmetry axis for deformed axially symmetric
nuclei. 
Following ideas of
Refs.~\cite{Bj74,BM75,Gr13,Gr19,Ju98} (see also Refs.~\cite{Ig83,So90}), 
we will use another
definition of the
spin-dependent level density $\rho_{\rm coll}(E,I)$ in terms of the
intrinsic level
density and collective rotation (and vibration) enhancement.
The level density
$\rho(E,\mathcal{K})$, e.g., 
Eqs.~(\ref{rho1}), (\ref{rho2}), and (\ref{rho2b}) for the level density
$\rho(E,M)$
at $M=\mathcal{K} $, is named in Ref.~\cite{Ig83} as an
intrinsic level density,
\be\l{rhointdef}
\rho_{\rm int}(U_{\rm coll},\mathcal{K})\equiv \rho(E,M=\mathcal{K}).
\ee
In Eqs.~(\ref{rho1}), (\ref{rho2}), (\ref{rho2b}) and
(\ref{rhointdef}), we replace 
$\Theta$ by the ``parallel'' moment of inertia $\Theta_{\parallel}$, and
$U_{\rm coll}$ by the excitation energy
$E-E_0-E_{\rm rot}$, where $E_{\rm rot}$ is the collective
rotation energy 
which depends explicitly
on  the spin
projection, $\mathcal{K}$, on the symmetry axis of an
axially-symmetric deformed nucleus.
Using these expressions, one can define the collective level density as \cite{Ig83}
\bea\l{collden}
&\rho_{\rm coll}^{}(E,I)=\frac{1}{2}\sum^{I}_{\mathcal{K}=-I}\rho_{\rm int}(U_{0}-E^{\perp}_{\rm rot},\mathcal{K})\\
  &\approx
  \frac{1}{2}\int_{-I}^{I} d \mathcal{K}~\rho_{\rm int}\left(U_{0}-E^{\perp}_{\rm rot},\mathcal{K}\right)~,
\eea
where $U_0=E-E_{0}$ and $E_{\rm rot}=E^{\perp}_{\rm rot}$.
The rotation energy $E^{\perp}_{\rm rot}$ for a collective
rotation around the axis $0x$ perpendicular to the symmetry
axis $0z$ \cite{BM75},
\be\l{ErotIK}
E^{\perp}_{\rm rot}=\left[I(I+1)-\mathcal{K}^2\right]/\left(2 \Theta_\perp\right),
\ee
where $\Theta_\perp$ is the corresponding (perpendicular to the symmetry axis) moment of inertia,
in contrast to the parallel MI $\Theta_{\parallel}$.
The factor 1/2 illuminates the reflection
degeneracy assuming that the deformation obey the reflection symmetry
with respect to the
plane, perpendicular to the symmetry axis $0z$. Shell
effects in the ``perpendicular'' moment of inertia, $\Theta_\perp$,
for any axially symmetric potential well are studied within the periodic
orbit theory \cite{MA02,MY11,BB03,SM76} in Ref.~\cite{GM21}.

Substituting Eqs.~(\ref{rho1}), (\ref{rho2}), or (\ref{rho2b})
with $\Theta=\Theta_\parallel$,
into Eq.~(\ref{collden}), one obtains the MMA expression for the level density,
$\rho_{\rm coll}^{}(E,I)$, in terms of the modified Bessel functions.
We expand now the expression (\ref{collden}) over the small spin
parameters, $E^\parallel_{\rm rot}/U_0$ and $E^{\perp}_{\rm rot}/U_0$.
Using the small-spin condition (\ref{Iexp}), taking approximately
$\Theta=\Theta_\perp$, and neglecting the effect of $\Theta=\Theta_\parallel$
in the exponent,
since $\Theta_\parallel\ll\Theta_\perp$, then one obtains,
\be\l{rhoIexp2coll}
  \rho_{\rm coll}^{}(E,I)=\rho^{\rm coll}_{\rm \tt{MMA}}(E,I) \approx
 \frac{\overline{\rho}^{}_{\nu}~ \mathcal{R}_{\rm coll}(I)~\exp\left(S_0\right)}{
4~S_0^{\nu-1}\sqrt{2\pi S_0}}~,
\ee
  %
where $S_0=2\sqrt{aU_0}$, and
\bea\l{qIexpcoll}
&\mathcal{R}_{\rm coll}(I)=\frac{2I+1}{q^2_\parallel}\sum_{\mathcal{K}=-I}^{I}
\exp\left(-\frac{I(I+1)}{2q_\perp^2}-\frac{K^2}{2 q^2_{\rm eff}}\right)\nonumber\\
&\approx \frac{(2I+1)^2}{q^2_\parallel}
\exp\left(-\frac{(I+1/2)^2)}{2q_\perp^2}\right)~.
\eea
We introduced here several dispersion parameters. Namely, one specifies
the effective dispersion parameter, $q^{2}_{\rm eff}$, the parallel,
$q^{2}_\parallel$, and perpendicular collective
    parameter, $q^{2}_\perp$, which are related to the
MI components, $\Theta_{\rm eff}$, $\Theta_{\parallel}$, and $\Theta_\perp$,
respectively,
\bea\l{qeff}
& q_{\rm eff}^2=\frac{\Theta_{\rm eff}}{\hbar^2}\sqrt{\frac{U_0}{a}}~,~~~
\frac{1}{\Theta_{\rm eff}}=\frac{1}{\Theta_{\rm \parallel}}+\frac{1}{\Theta_{\rm \perp}}~,\\
&q_{\rm \parallel}^2=\frac{\Theta_{\rm \parallel}}{\hbar^2}\sqrt{\frac{U_0}{a}}, ~~
q_{\rm \perp}^2=\frac{\Theta_{\rm \perp}}{\hbar^2}\sqrt{\frac{U_0}{a}}~.
\eea
The last term in the exponent argument in the first expression of
Eq.~(\ref{qIexpcoll})
can be neglected for $K^2\ll q^2_{\rm eff}$.
As seen from comparison of Eq.~(\ref{qIexpcoll}) with
Eq.~(\ref{qIexp}),
the collective rotation
is much enhanced by the perpendicular dispersion parameter,
$q^2_{\perp}\gg q^2_{\parallel}$,
in line of the results obtained in
Refs.~\cite{Bj74,BM75,Gr13,Gr19,Ju98}.

A.S.~Davydov and his collaborators \cite{DF58,DC60}
have introduced in nuclear physics the possible existence
of non-axially symmetric ground-state deformations in some nuclei
and applied their model to the description of collective rotations.
For non-axial collective rotations, one can obtain the expression for the level 
density in terms of the
internal level density, $\rho_{\rm int}(U_0-E^{\rm nax}_{\rm rot})\equiv
\rho_{\rm int}(U_{0}-E^{\rm nax}_{\rm rot},0)$,
modified with the
non-axially rotational energy, $E^{\rm nax}_{\rm rot}$,
in terms of the modified Bessel functions.
Expanding over small spin parameters, $E^{\rm nax}_{\rm rot}/U_0$
[Eq.~(\ref{Iexp})], one finds
\cite{Ig83} 
%
\be\l{levdenna}
\rho_{\rm coll}^{}(E,I)
\approx \frac{1}{2}(2I+1)
\rho_{\rm int}\left(U_0\right)
~\exp\left[-\frac{(I+1/2)^2}{2\overline{q^2}}\right]~,
\ee
where $\overline{q^2}=(q_x^2+q_y^2+q_z^2)/3$ is the
averaged value of the spin dispersion
    parameter, $q_j^2=\Theta_{j}\sqrt{U_0/a}/\hbar^2$
    are partial dispersions, $j=x,y$, and $z$.
    In these derivations, we neglected  partial
$\mathcal{K}^2$ components with respect to $\overline{q^2}$, and
$I(I+1)\approx (I+1/2)^2$ in the semiclassical approximation.

\subsection{Total integrated MMA level densities}
\l{subsec3-3}

For a comparison with experimental data
\cite{ENSDFdatabase}, we present also the level density $\rho(E,N,Z)$
obtained by the integration of $\rho(E,N,Z,M)$ over the angular momentum
projection $M$. The statistics conditions have improved \cite{PRC,IJMPE},
so that we can study a longer chain of isotopes, including those relatively far
away from the beta stability line.
Including
the shell and isotopic asymmetry effects, for the standard saddle-point method
approach [Eq.~(\ref{SPMgennp})]
      one arrives at (see Ref.~\cite{IJMPE})
\bea\l{SPMgennptotdef}
&\rho(E,N,Z) = \sum_M \rho(E,N,Z,M)\approx \int d M \rho(E,N,Z,M)\\
&\approx
\frac{a^{3/4}\exp(2\sqrt{aU_0})}{4\pi~U^{5/4}~\sqrt{\pi\mathcal{J}_0(1+\xi^\ast)}}
\approx
\frac{\sqrt{\pi}\exp(2\sqrt{aU_0})}{12 a^{1/4}~U^{5/4}~\sqrt{1+\xi^\ast}}~.
\l{SPMgennptot}
\eea
Here, the summation and approximate integration are carried out over all spin
projections $M$. In Eq.~(\ref{SPMgennptot}), $a=\pi^2 g(\lambda)/6$
is the total level density parameter,
    Eqs.~(\ref{anp}) 
and (\ref{daFnp}); $\mathcal{J}_0$ is the component of the
Jacobian $\mathcal{J}$,
Eq.~(\ref{J12np}), which is independent of
$\beta$ but depends on the shell structure; $U_0$ is the 
excitation energy
$U$ [Eq.~(\ref{Eex})] at zero angular momentum projection, $M=0$;
see also Eq.~(\ref{pars})
for $\xi^\ast$.
%
%
 The asterisk means $\beta=\beta^\ast=\sqrt{a/U}$ at the saddle point.
 In the second equation of (\ref{SPMgennp}), and of (\ref{pars})
 we also used
    $\lambda_n\approx \lambda_p\approx \lambda$
 for a small asymmetry parameter, $X^2$, 
 together with Eq.~(\ref{d2Edl2}) for the
derivatives of the energy shell corrections 
    and Eq.~(\ref{periode}) for
the mean distance between
neighboring major shells near the Fermi surface,
$\mathcal{D}\approx \lambda/A^{1/3}$.


    As in section \ref{subsec2-2}, 
    at large excitation energy, Eq.~(\ref{SPMgennptot})
     is a more general shell-structure Fermi-gas (SFG) 
    asymptotic
   with respect to the well-known \cite{Be36,Er60,GC65}
   Fermi gas (FG) 
  approximation for $\rho(E,N,Z)$, which is equal to Eq.~(\ref{SPMgennptot}) at
  $\xi^\ast \rightarrow 0$,
  \be\l{FGtot}
\rho(E,N,Z)\rightarrow 
\frac{\sqrt{\pi}\exp(2\sqrt{aU})}{12 a^{1/4}~U^{5/4}}~.
\ee

Similarly, as in section \ref{subsec2-3}, according to the general definitions in
Eq.~(\ref{SPMgennptotdef}),
for the total integrated MMA
level density, one has Eq.~(\ref{denbesnp}) (Ref.~\cite{IJMPE}) but
    with the entropy $S$
    given by $S=2\sqrt{a U_0}$ [Eq.~(\ref{Eex}) for the excitation energy $U$ but at $M=0$]. 
     In cases (i) and (ii),
     named below the MMA1 and MMA2 approaches,  respectively,
    one obtains
    Eq.~(\ref{denbesnp})
    with different coefficients $\overline{\rho}_\nu$ 
    (see also Refs.~\cite{PRC,IJMPE}),
\bea\l{rho1tot}
&\rho^{}_{\tt{MMA1}}(S)=\overline{\rho}_{2}S^{-2}I_{2}(S),\qquad
\overline{\rho}_{2}=\frac{2 a^2}{\pi\sqrt{\mathcal{J}_0}}\approx \frac{2\pi a}{3}
\qquad \mbox{(i)},\\
&\rho^{}_{\tt{MMA2}}(S)=\overline{\rho}_{3}S^{-3}I_{3}(S),\qquad
\overline{\rho}_{3}\approx 
\frac{4a^3}{\pi \sqrt{\overline{\xi}\mathcal{J}_0}}
\approx \frac{4\pi a^2}{3\sqrt{\overline{\xi}}} \qquad \mbox{(ii)},
\l{rho2tot}~
\eea
with the same $\overline{\xi}$ and $\mathcal{J}_0$
given by
Eqs.~(\ref{xiparnp}) and
(\ref{J12np}), respectively; see also Eq.~(\ref{J02snp}).
For the TF approximation to the coefficient $\overline{\rho}_{3}$ within the
 case (ii), 
one finds \cite{NPA,PRC,IJMPE}
\be\l{rho2btot}
\rho^{}_{\tt{MMA2b}}(S)=\overline{\rho}^{(2b)}_{3}S^{-3}I_{3}(S),\qquad
\overline{\rho}^{(2b)}_{3}\approx \frac{2 \sqrt{6}~\lambda a^2}{3}~.
\ee

\section{Discussion of the results}
\label{sec4}

  Fig.~\ref{fig2} and Table I 
show results of
different theoretical approaches
[MMA, Eqs.~(\ref{rho1tot}), (\ref{rho2tot}), and (\ref{rho2btot});  SFG,
  Eq.~(\ref{SPMgennptot}), 
and;  standard FG, Eq.~(\ref{FGtot})]
for the statistical level density $\rho(E,N,Z)$ (in logarithms)
as functions of the
excitation energy $U$. 
They are compared to the
experimental
data obtained by the sample method \cite{So90,PRC,IJMPE}.
In Fig.~\ref{fig2} and Table I, 
we present  
the results of the total MMA level density, $\rho(E,N,Z)$,
[Eqs.~(\ref{rho1tot}) for the MMA1, (\ref{rho2tot}) for the MMA2 and
  (\ref{rho2btot}) for the MMA2b approaches] for the inverse
    level-density
parameter $K$ obtained from a least mean-square 
fit of the calculated level density to the
experimental data \cite{ENSDFdatabase} that was deduced by
using the sample method \cite{PRC,IJMPE,NPA}.
The control relative error-dispersion parameter $\sigma$
is determined in terms of $\chi^2 $ of the least mean-square
fit by the standard formulas:
\begin{equation}\label{chi}
\sigma^2=\frac{\chi^2}{\aleph-1}, \qquad \chi^2 =\sum_i
\frac{(y(U_i)-y^{\rm exp}_i)^2}{(\Delta y_i)^2},\qquad y=\ln\rho~,
  \end{equation}
where $\aleph$ is the sample number and $\Delta y_i \sim 1/\sqrt{N_i}$
and $N_i$ is a number of level states in the sample,
$i=1,2,3\ldots,\aleph$.

\begin{turnpage}
  \begin{table}
    \caption{
      The
  inverse level-density parameter $K$  (with errors $\Delta K$ 
  in parenthesis) in units of MeV,
   found by the  least mean-square 
   fit  for 
      $^{131-156}$Nd in the low energy states ranges
      restricted by maximal values of the excitation energy
      having clear spins 
      (from the ENSDF database \cite{ENSDFdatabase}),
      and $U_{\rm max}$ (also in MeV units), with the
    precision of the standard expression for $\sigma$, Eq.~(\ref{chi}),
    are shown for 
  several approximations 
  with the same notations as in Fig.~\ref{fig2}; 
  asterisks mean results of calculations, accounting for pairing
  with $E_{\rm cond}=0.72$ MeV for $A=140$ and $0.69$ MeV for $A=142$;
  see also text. The MMA approaches are presented with 
       minimal $\sigma$ which 
     were  obtained
          for 
     the corresponding one-component 
     systems of $A$ nucleons \cite{PRC}. 
     \label{table}
     }
{
\begin{tabular}{@{}ccccccccccccccc@{}}\toprule
    & &
\multicolumn{2}{c}{FG}    &
\multicolumn{2}{c}{SFG}   &
\multicolumn{2}{c}{MMA1}  &
\multicolumn{2}{c}{MMA2a} &
\multicolumn{2}{c}{MMA2b} &
\multicolumn{3}{c}{One-component system} \\ \cline{3-12}
    &
\multirow{3}{*}{\minitab[c]{$U_{\rm max}$\\ MeV}} &
\multirow{3}{*}{\minitab[c]{$K$($\Delta K$)\\ MeV}} & &
\multirow{3}{*}{\minitab[c]{$K$($\Delta K$)\\ MeV}} & &
\multirow{3}{*}{\minitab[c]{$K$($\Delta K$)\\ MeV}} & &
\multirow{3}{*}{\minitab[c]{$K$($\Delta K$)\\ MeV}} & &
\multirow{3}{*}{\minitab[c]{$K$($\Delta K$)\\ MeV}} & & &
\multirow{3}{*}{\minitab[c]{$K$($\Delta K$)\\ MeV}} & \\
$A$ & & & $\sigma$ & & $\sigma$ & & $\sigma$ & & $\sigma$ & & $\sigma$ &
 Approach & & $\sigma$ \\[-1ex]
 &&&&&&&&&&&&&& \\ \colrule
131 & 1.57 &
7.56~(1.2) & 10.8 &  
7.47~(1.1) & 19.7 &  
7.01~(1.1) & 12.3 &  
8.49~(0.92) & 9.1 &  
18.8~(1.4) & 4.5 &  
MMA2b &
26.2~(1.6) & 3.7 \\
132 & 3.66 &
18.7~(1.0) & 4.4 &  
18.3~(0.9) & 4.3 &  
17.7~(1.1) & 5.4 &  
19.4~(0.7) & 3.5 &  
34.9~(0.3) & 0.7 &  
MMA2b &
47.2~(0.6) & 0.8 \\
133 & 0.55 &
~3.8~(0.2) & 2.2 &  
~3.8~(0.2) & 2.2 &  
~3.5~(0.2) & 2.7 &  
~7.3~(0.2) & 1.0 &  
13.2~(0.5) & 1.1 &  
MMA2b &
19.0~(0.8) & 1.3 \\
134 & 1.42 &
10.7~(0.7) & 1.7 &  
10.2~(0.6) & 1.7 &  
9.89~(0.7) & 2.1 &  
11.3~(0.5) & 1.6 &  
26.0~(1.5) & 1.3 &  
MMA2b &
39.1~(3.8) & 2.2 \\
135 & 0.56 &
~4.5~(0.6) & 3.2 &  
~4.5~(0.6) & 3.2 &  
~4.1~(0.6) & 3.6 &  
~10.0~(0.8) & 2.0 &  
15.8~(1.3) & 1.7 &  
MMA2b &
23.6~(2.4) & 2.0 \\
136 & 1.75 &
16.6~(1.1) & 1.4 &  
15.2~(0.9) & 1.3 &  
14.9~(1.2) & 1.8 &  
25.6~(0.9) & 0.8 &  
39.4~(1.1) & 0.5 &  
MMA2b &
56.4~(1.6) & 0.5 \\
137 & 2.47 &
~13.7~(1.1) & 5.7 &  
~13.7~(1.1) & 5.7 &  
~12.8~(1.1) & 6.9 &  
~21.8~(0.8) & 2.7 &  
28.4~(0.8) & 2.0 &  
MMA2b &
38.9~(1.1) & 1.8 \\
138 & 2.32 &
15.9~(0.7) & 1.9 &  
15.8~(0.7) & 1.9 &  
15.0~(0.5) & 1.7 &  
19.0~(0.9) & 2.3 &  
32.8~(2.6) & 3.2 &  
MMA2a &
23.0~(1.2) & 2.5 \\
139 & 0.93 &
~7.3~(0.9) & 3.3 &  
~7.2~(0.9) & 3.3 &  
~6.7~(0.9) & 3.8 &  
~9.5~(0.8) & 2.7 &  
21.0~(1.8) & 2.0 &  
MMA2b &
32.7~(3.6) & 2.3 \\
140 & 3.49 &
19.1~(0.5) & 2.3 &  
18.3~(0.5) & 2.2 &  
18.2~(0.5) & 2.3 &  
18.7~(0.4) & 2.3 &  
35.0~(1.5) & 3.2 &  
MMA2a &
22.4~(0.6) & 2.3 \\
& & & & & & & & & & & & MMA2b* &30.3~(1.4) & 2.7\\
%
141 & 1.62 &
~12.7~(0.9) & 2.1 &  
~12.0~(0.8) & 2.1 &  
~11.8~(0.9) & 2.3 &  
~13.0~(0.7) & 2.0 &  
29.3~(1.8) & 1.7 &  
MMA2b &
42.6~(4.0) & 2.3 \\
142 & 3.42 &
19.8~(0.4) & 1.3 &  
17.7~(0.3) & 1.4 &  
18.9~(0.3) & 1.1 &  
16.9~(0.2) & 1.2 &  
36.1~(1.4) & 2.6 &  
MMA2a &
20.5~(0.3) & 1.4 \\
& & & & & & & & & & & & MMA2b* &31.5~(1.3) & 2.2\\
%
143 & 1.91 &
~13.5~(0.7) & 2.1 &  
~12.6~(0.6) & 2.1 &  
~12.7~(0.7) & 2.2 &  
~13.1~(0.6) & 2.0 &  
29.4~(1.3) & 1.5 &  
MMA2b &
43.2~(2.4) & 1.8 \\
144 & 2.30 &
17.0~(0.7) & 1.7&  
16.0~(0.7) & 1.7 &  
16.0~(0.7) & 1.7 &  
15.8~(0.6) & 1.7 &  
35.3~(2.3) & 2.3 &  
MMA2a &
19.2~(1.0) & 2.3 \\
145 & 1.72 &
~10.5 (0.5) & 2.9 &  
~10.3~(0.4) & 2.9 &  
~9.9~(0.6) & 4.1 &  
~11.1~(0.4) & 2.7 &  
24.0~(0.7) & 1.7 &  
MMA2b &
33.1~(1.2) & 2.0 \\
146 & 1.81 &
14.0 (0.7) & 1.7 &  
13.6~(0.6) & 1.8 &  
13.2~(0.6) & 1.8 &  
14.1~(0.6) & 1.8 &  
31.3~(2.4) & 2.5 &  
MMA2a &
17.0 (0.7) & 1.8 \\
147 & 1.21 &
~8.3~(1.0) & 5.6 &  
~9.1~(0.8) & 4.9 &  
~7.6~(1.0) & 6.9 &  
~8.2~(0.9) & 5.6 &  
21.7~(1.3) & 2.5 &  
MMA2b &
3.6~(1.8) & 2.2 \\
148 & 1.78 &
12.1 (0.5) & 2.1 &  
15.5 (0.8) & 2.6 &  
11.4 (0.4) & 2.2 &  
12.2 (0.4) & 2.2 &  
26.9~(1.9) & 2.3 &  
MMA2a &
14.7~(0.6) & 2.3 \\
149 & 0.81 &
~5.1~(0.7) & 6.6 &  
~5.1~(0.7) & 6.5 &  
~4.7~(0.5) & 4.9 &  
~6.3~(0.6) & 5.3 &  
16.5~(1.1) & 2.7 &  
MMA2b &
23.6~(1.4) & 2.3 \\
150 & 1.20 &
~9.5~(0.6) & 2.1 &  
~9.3~(0.5) & 2.1 &  
~8.9~(0.6) & 2.6 &  
~9.8~(0.5) & 2.1 &  
24.6~(1.9) & 2.4 &  
MMA2a &
12.1~(0.4) & 1.4 \\
151 & 1.93 &
~3.9~(0.8) & 7.2 &  
~3.5~(0.7) & 8.5 &  
~3.5~(0.7) & 8.5 &  
~4.8~(0.6) & 5.9 &  
14.5~(1.4) & 3.1 &  
MMA2b &
21.0~(1.9) & 2.8 \\
152 & 1.90 &
12.4~(0.7) & 3.4 &  
11.8~(0.6) & 3.3 &  
11.7~(0.8) & 4.3 &  
11.7~(0.5) & 3.5 &  
27.6~(1.7) & 3.4 &  
MMA2a~ &
14.0~(0.6) & 3.3 \\
153 & 1.58 &
~9.2~(1.3) & 9.6 &  
~8.9~(1.2) & 9.5 &  
~8.5~(1.3) & 11.3 &  
~9.1~(1.0) & 9.0 &  
22.7~(1.4) & 3.7 &  
MMA2b &
31.7~(1.7) & 3.0 \\
154 & 1.35 &
10.9~(0.8) & 2.6 &  
10.5~(0.7) & 2.5 &  
10.1~(0.9) & 3.4 &  
10.6~(0.6) & 2.6 &  
27.8~(1.5) & 1.8 &  
MMA2a &
13.0~(0.8) & 2.7 \\
155 & 1.83 &
13.7~(2.4) & 6.0 &  
12.9~(2.1) & 5.9 &  
12.2~(2.4) & 7.7 &  
12.1~(1.7) & 6.4 &  
32.2~(2.5) & 2.6 &  
MMA2b &
47.8~(5.3) & 3.5 \\
156 & 2.74 &
17.6~(1.5) & 6.0 &  
16.1~(1.2) & 5.8 &  
16.5~(1.6) & 7.2 &  
15.4~(1.1) & 6.4 &  
35,7~(1.6) & 2.9 &  
MMA2b &
48.9~(2.2) & 2.7  \\\botrule
\end{tabular}
}
\end{table}
\end{turnpage}

\begin{figure}
   \centerline{\includegraphics*[width=8cm,clip]{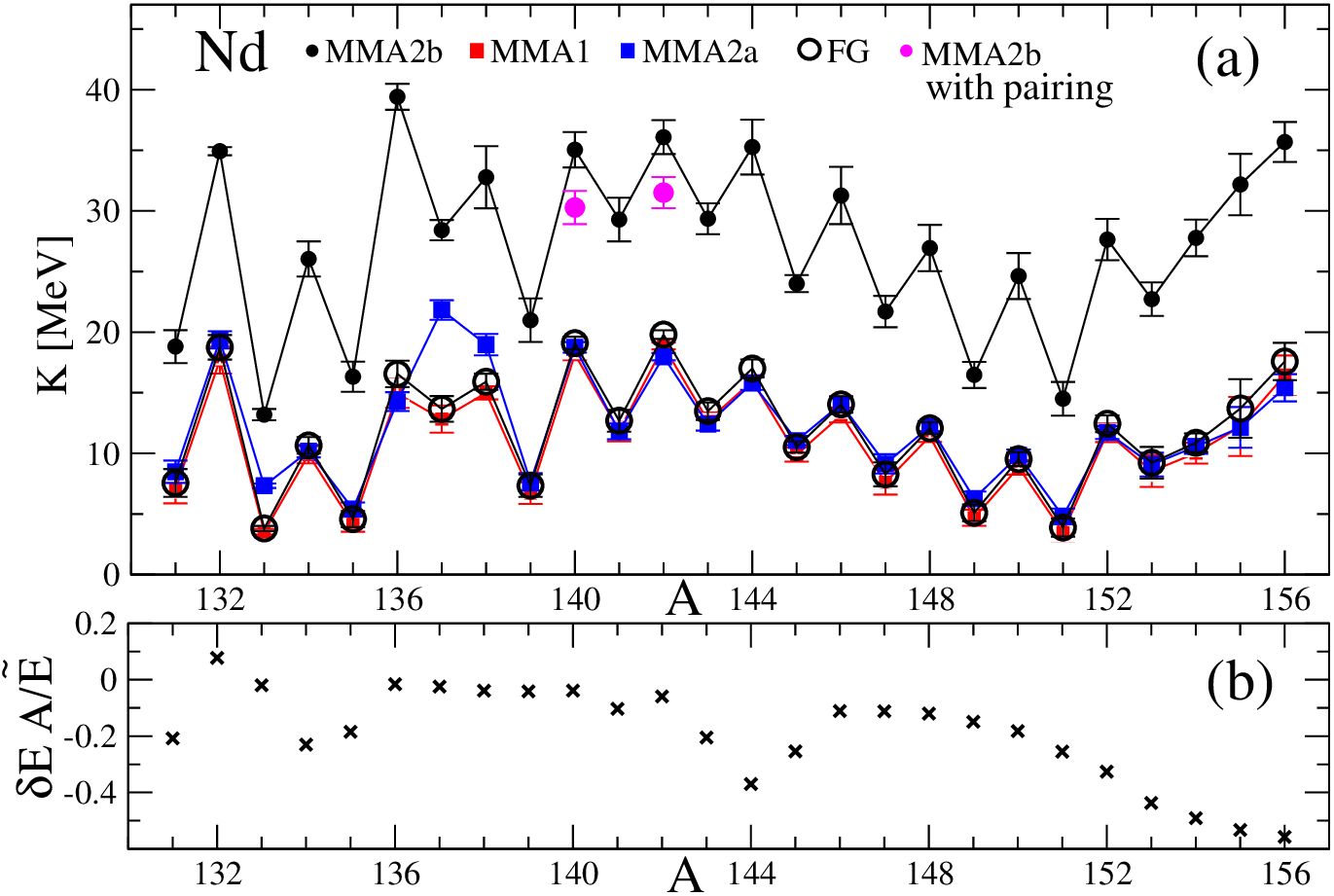}}
   \caption{
      Inverse level-density parameters $K$ (with errors) 
      for available Nd isotopes 
      are shown as function of the particle numbers $A$ within a
     long chain $A=131-156$ [panel (a)].
     Different symbols correspond to several approximations:
      the close black
     dots for the MMA2b [Eq.~(\ref{rho2btot})],
     full red squares for MMA1 [Eq.~(\ref{rho1tot})],
     larger blue open squares for MMA2a [Eq.~(\ref{rho2tot})],  and
     big open circles 
     for Fermi gas (FG) [Eq.~(\ref{FGtot})] approaches.
     ``MMA2b with pairing'' points account for the excitation
     energy shift due to pairing condensation \cite{Ig83,So90,PRC,SC19}
     with the best (smallest) $\sigma$ compared with other approaches.
     In panel (b),
     the relative energy shell corrections, $\delta E$, are shown in
     units of the averaged energy per particle number $A$ \cite{MSIS12}.
}
\label{fig3}
\end{figure}

\begin{figure}
  \centerline{\includegraphics*[width=8cm]{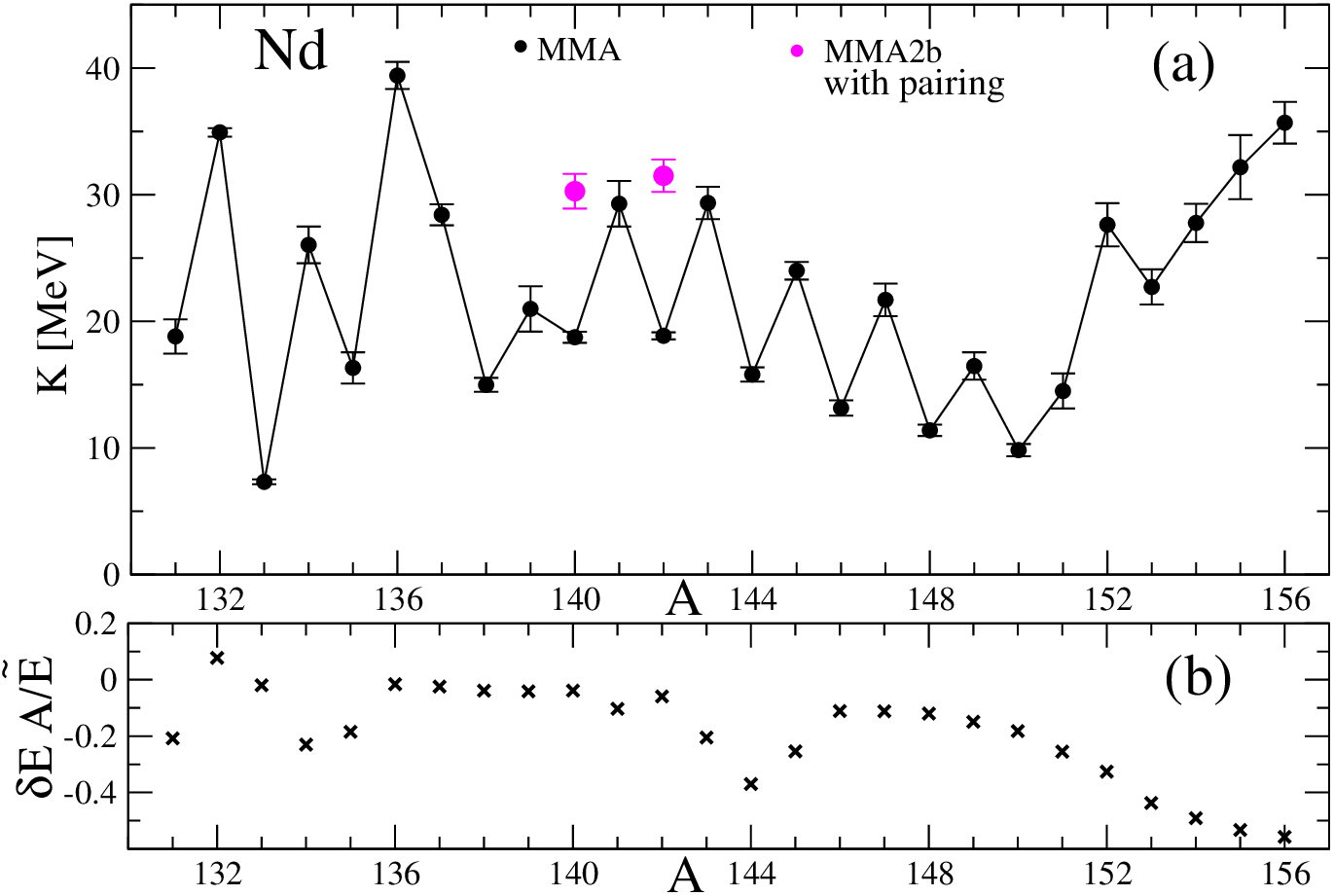}}
  \caption{
      (a) Inverse level-density parameter $K$ (with errors) 
      for Nd isotopes 
      is shown as
     function of the particle number $A$ within a long chain A=131-156:
     the close black 
     dotted points are 
     for the MMA approach
      taken with the smallest relative 
     error parameter
    $\sigma$, Eq.~(\ref{chi}), among all MMAs. (b):  The
         relative shell correction energies,
         $\delta E A/\tilde{E}$, are taken
         from Ref.~\cite{MSIS12}.
}
\label{fig4}
\end{figure}

   We determine $\sigma$, Eq.~(\ref{chi}), at the minimum of
    $\chi^2$ over the unique parameter, $K=K_{\rm min}$,
    having a definite physical meaning as the
    inverse level-density parameter $K$
    (see Figs.~\ref{fig2} and \ref{fig3}).
      Then, we may compare the values of $\sigma$ for
    several different  MMA approximations to the level density
    approaches (Fig.~\ref{fig3}), which were found independently of 
    the data,
  under certain statistical conditions mentioned above.
    For this aim we are interested in the lowest value obtained for
  $\sigma(K_{\rm min})$ by fitting calculated results of different theoretical
  approaches. The MMA  results for the minimal values
  of $\sigma$,
  Eq.~(\ref{chi}), are shown
  in plots of Fig.~\ref{fig4} by black solid lines as the best, among the
  MMA approaches, 
  agree with the
  experimental data.  The results of our calculations
    are almost independent of
    the sample number, $\aleph=5-7$, which plays the same role as an
    averaging parameter on
    the plateau condition in the Strutinsky averaging procedure
    \cite{BD72}.

    As done in Refs.~\cite{PRC,IJMPE} for
    several isotopes, 
    Fig.~\ref{fig2}
and Table I   
 present 
the two opposite situations
concerning the states distributions as functions of the excitation energy
$U=U_0=E-E_0$ [Eq.~(\ref{Eex}) at $M=0$].
We 
show results for the nucleus $^{145}$Nd (b)
with a large number of the low energy states 
below excitation energy of about
1 MeV. 
For $^{142}$Nd (a)
one has 
a very small number of low energy states 
below 
    the same energy of about 1 MeV 
 (see ENSDF database \cite{ENSDFdatabase} 
    and Table I 
    for maximal excitation energies
 $U_{\rm max}$).
But there are many states in $^{142}$Nd with 
 excited energies of
above 1 MeV up to essentially 
larger
    excitation energy of about 2-4 MeV. 
    According to Ref.~\cite{MSIS12},
the shell effects, measured by
$\mathcal{E}_{\rm sh}$, Eq.~(\ref{xibdEnp}), are significant in both
these nuclei, a slightly deformed $^{145}$Nd
and spherical $^{142}$Nd, see Fig.~\ref{fig3}(b).

In 
Fig.~\ref{fig2} and Table I, 
the results of the 
MMA1 and MMA2 approaches, 
Eqs.~(\ref{rho1tot}) and (\ref{rho2tot}), respectively,
are compared with the  FG approach, Eq.~(\ref{FGtot}).
The SFG results, Eq.~(\ref{SPMgennptot}), are
very close
    to those of the well-known FG asymptote, Eq.~(\ref{FGtot}),
     which neglects
the shell effects; see Table I. 
Therefore, they are not shown in Fig.~\ref{fig2}.
The results of the MMA2a approach, Eq.~(\ref{rho2tot}), 
in the dominating 
    shell effects case (ii)
    [$\xi^\ast \gg 1$, Eq.~(\ref{pars})]
with 
the realistic relative shell correction,
$~\mathcal{E}_{\rm sh}~$ (Ref.~\cite{MSIS12}),
 are shown 
versus those of  a  small 
shell effects approach MMA1 (i), 
 Eq.~(\ref{rho1tot}), valid at
        $~\xi^\ast \ll 1~$. 
The results of the
limit of the MMA2 approach to a very small 
value of $~\mathcal{E}_{\rm sh}$, but still within the case (ii),
Eq.~(\ref{rho2btot}),
named as MMA2b, 
are also shown in Fig.~\ref{fig2} and Table I 
because of a large shell
structure contribution due to relatively large derivatives
of the energy shell corrections
over the chemical potential. They are
in contrast to
the results of the
MMA1 approach.
 The results of the
 SFG asymptotical full saddle-point 
 approach, 
Eq.~(\ref{SPMgennptot}), and of a similar popular FG approximation,
Eq.~(\ref{FGtot}), are both in good agreement with those of the
standard Bethe formula \cite{Be36} for 
one-component systems 
    (see Ref.~\cite{PRC}), and are also presented 
in Table I. 
      For finite realistic values of
      $\mathcal{E}_{\rm sh}$,  the value of the inverse level-density
      parameter $K$ of the MMA2a  (Table I) 
      and the
      corresponding level density
      (Fig.~\ref{fig2})
      are in between those of the MMA1 and MMA2b. 
      Sometimes, the results of the MMA2a approach 
       are significantly
      closer to those of the MMA1 
      one, than to those of the MMA2b approach, e.g., for nuclei
      as $^{142}$Nd.
           
In both panels of 
Fig.~\ref{fig2}, one can see the divergence
of the FG approach [Eq.~(\ref{FGtot})], as well
as of the SFG approach [Eq.~(\ref{SPMgennptot})], in the
full SPM level-density asymptote  in
the zero-excitation energy limit $U\rightarrow 0$. 
    This is clearly seen also analytically, in particular  
in the FG limit,  Eq.~(\ref{FGtot}); 
see also the general asymptotic  expression
(\ref{rhoasgennp}). 
It is, obviously, in contrast to any MMAs
combinatorics
expressions 
(\ref{den0gennp})
in this limit;
see Eqs.~(\ref{rho1tot})-(\ref{rho2btot}).
The MMA1 results are close to those of the 
 FG and SFG approaches 
    for all considered nuclei
    (Table I), 
    in particular,
for both 
$^{142}$Nd and $^{145}$Nd isotopes 
in Table I. 
The reason is that
their differences are
essential only for extremely small excitation energies $U$, where
 the MMA1 approach is finite while 
other,  FG and SFG,
approaches 
are divergent. 
However, there 
are almost no experimental data for 
 excited states
in the range of their differences, at least in the nuclei under
consideration.

    The MMA2(b) results,
  Eq.~(\ref{rho2btot}), 
  for
$^{145}$Nd [see Fig.~\ref{fig2}(b)] with $\sigma \sim 1$
   are 
significantly better in agreement with the experimental
data
as compared 
to the results of all other approaches
(for the same nucleus).
 For this nucleus, the MMA1 [Eq.~(\ref{rho1tot})],
FG [Eq.~(\ref{FGtot})],
    and 
    SFG [Eq.~(\ref{SPMgennptot})] 
approximations 
    are characterized by
    much larger $\sigma$  
     (see Table I). 
    In contrast to the case of 
    $^{145}$Nd [Fig.~\ref{fig2}(b)] with
    excitation energy spectrum 
     having a 
     large number of low energy states 
     below
    about 1 MeV, for
    $^{142}$Nd [Fig.~\ref{fig2}(a)]  with almost no such states  in the same
     energy range, one 
    finds the opposite case~--~
    a significantly 
     larger MMA2b value of $\sigma$ 
     as compared to those
     for
    other approximations 
    (Fig.~\ref{fig2} and Table I). 
    In
    particular, for  MMA1 [case (i)], and
    other asymptotic approaches
     FG and SFG, 
   one obtains  for $^{142}$Nd spectrum almost the same 
$\sigma \sim 1$,
and almost the same for 
MMA2a [case (ii)] with realistic values of
$\mathcal{E}_{\rm sh}$. 
 Again, notice 
that 
the MMA2a results [Eq.~(\ref{rho2tot})]
 are closer, at the realistic $\mathcal{E}_{\rm sh}$,
to  those of the MMA1 
[case (i)], as well as
 the results of the  FG and SFG approaches.
The MMA1 and MMA2a results (at
realistic values of $\mathcal{E}_{\rm sh}$) as well as 
 those of the
 FG and SFG approaches 
are
obviously better in agreement with the
experimental data \cite{ENSDFdatabase} (see Refs.~\cite{PRC,IJMPE}) 
for $^{142}$Nd [Fig.~\ref{fig2}(a)].

One of the reason for the exclusive properties of $^{145}$Nd 
 [Fig.~\ref{fig2}(b)],
    as compared to $^{142}$Nd [Fig.~\ref{fig2}(a)],
        might be assumed to be 
        the nature
    of the excitation energy
    in these nuclei.
    Our MMAs results [case (i)] or [case (ii)]
    could clarify the excitation nature
        as 
    assumed in Refs.~\cite{PRC,IJMPE}.  
   Since 
    the  MMA2b results [case (ii)] are 
    much better
    in agreement with the experimental data than the MMA1 results
    [case (i)]
    for $^{145}$Nd,
    one could presumably 
    conclude
    that for $^{145}$Nd 
    one finds more
    clear thermal low-energy excitations. 
    In contrast to this, for   $^{142}$Nd
    [Fig.~\ref{fig2}(a)], one observes more regular high-energy
    excitations due to, e.g.,  
    the dominating rotational excitations,
            see Refs.~\cite{KM79,PRC,IJMPE}.
   As seen, in particular, from the values of the inverse 
       level-density parameter $K$ and the shell structure of
   the critical quantity,
Eq.~(\ref{pars}),
these properties can be 
understood 
to be mainly due to the larger values of
$K$ and shell correction second derivative $\mathcal{E}^{\prime\prime}_{\rm sh}$,
for low energy states 
in  $^{145}$Nd
 (Table I) 
 versus those of the $^{142}$Nd spectrum. 
 This is  in addition to the shell effects, which 
    are very important 
    for the case (ii) 
     which is not even realized 
    without    
  their dominance.

  As results, the statistically averaged level densities $\rho(E,N,Z)$
  for the MMA approach with a minimal value of the control-error
  parameter $\sigma$, Eq.~(\ref{chi}), in plots of
 Fig.~\ref{fig2} 
 agree well with those of the experimental data.
 The results of the MMA,  SFG and FG approaches
 for the level densities
 $\rho(E,N,Z)$ 
 in
Fig.~\ref{fig2}, and for $K$ in
Table I, 
do not depend 
on the cut-off spin factor
and moment of inertia 
because of the summations (integrations) over all spins projections,
or over spins, 
indeed, with accounting for the degeneracy $2I+1$ factor.
We do not use 
empiric free fitting parameters in our calculations, in particular, for
the FG results shown  in 
Table I, 
in contrast
to the back-shifted Fermi gas \cite{DS73}
and constant temperature models, see also
    Ref.~\cite{EB09}.

The results of calculations 
for the inverse
level-density parameter $K$ 
in the long Nd isotope chain with $A=131-156$ are summarized in
Fig.~\ref{fig3}  and Table I. 
 Preliminary spectra data for nuclei far away from the
    $\beta$-stability line
     from
     Ref.~\cite{ENSDFdatabase} are included in comparison with
     the results of the 
     theoretical approximations. These experimental data may
    be incomplete.
    Nevertheless, 
    it might be helpful to present 
    a comparison between theory and experiment to check
    general common effects of the 
    isotopic asymmetry and shell structure
    in a wide
    range of nuclei around
    the $\beta$-stability line. 
    
    As seen in Fig.~\ref{fig3},
        the results for $K$ for the isotopes of Nd
    ($Z=60$) as a function of 
    the particle number $A$  
    are characterized by a very pronounced saw-toothed behavior
    with alternating
low and high $K$ values for odd and even nuclei, respectively.
As for the platinum chain in Ref.~\cite{IJMPE}, this behavior is more pronounced
for the MMA2b (close black dots) approach with larger $K$ values.
 For each nucleus, the significantly
smaller MMA1 value of $K$ (full red squares) is 
close to  that of 
the FG approach (heavy open black circles).
The SFG results are 
    very close to those of the FG approach and,
therefore, 
are not shown
in the plots, but presented in Table I. 
The MMA2a results for $K$ are intermediate between the MMA2b and MMA1 ones,
 but closer to the MMA1 values.

Notice that for the rather
long chain of isotopes of Nd, as for Pt ones \cite{IJMPE},
    one finds a remarkable shell oscillation
    (Fig.~\ref{fig3}).
Fixing the even-even (even-odd) chain, for all compared approximations,
one can see a hint of slow oscillations by evaluating 
its period  $\Delta A$,
$\Delta A \sim 
 30$ for $A\sim 140$
(see Refs.~\cite{NPA,IJMPE}).
  Within order of magnitude,  these estimates agree
 with the
 main period for the relative shell corrections,
 $\delta E A/\tilde{E}$,
shown in Fig.~\ref{fig3}(b). 
Therefore,
according to these evaluations and Fig.~\ref{fig3}(b)
for $K(A)$,  we show the sub-shell effects within a major shell. 
This shell oscillation as function of $A$ is
more pronounced for the MMA2b case because of its relatively large amplitude,
but is mainly proportional to that of the MMA2a and other approximations.

The MMAs results shown in Fig.~\ref{fig3}, as function of
    the
    particle number $A$, Eq.~(\ref{SPMgennptot}),
 can be partially understood  through the basic 
 critical quantity, $\xi\sim \xi^\ast \leq  \xi^\ast_{\rm max}$, where
     $ \xi^\ast_{\rm max} 
    \propto
       KU_{\rm max}\mathcal{E}^{\prime\prime}_{\rm sh}/A^{4/3}$. 
       Here we need also  the maximal
      excitation energies
      $U_{\rm max}$ of the low energy states
      (from Ref.~\cite{ENSDFdatabase} and Table I) 
      used in our calculations. Such
      low energy states  
      spectra are more 
    complete due to information on
        the spins 
   of the states. 
     As assumed in the derivations
            (Subsection~\ref{subsec2-3}), larger values of
    $\xi^\ast$, Eq.~(\ref{pars}), are 
   expected in the MMA2b
   approximation (see Fig.~\ref{fig3}),
  due to large values of $K$ (small level density parameter $a$).
   For the MMA1 approach,
     one finds significantly smaller $\xi^\ast$, and 
         in between values (closer to those of the MMA1)
     for the MMA2a case.
     This is  in line 
    with the assumptions for case (i) and case (ii)
     in the derivations of the level-density approximations
of the MMA1, Eq.~(\ref{rho1}),
    and MMA2, Eq.~(\ref{rho2}), respectively.

    In order to clarify the
    shell effects, we present in Fig.~\ref{fig4}(a)
    the inverse level-density parameters $K(A)$ 
     taking the MMA
        results with the smallest values of $\sigma$
      for each nucleus (see also Fig.~\ref{fig3} and Table I). 
     Among all MMAs results, this provides  
     the best agreement 
     with
     the experimental data for the statistical level density obtained
     by the sample method \cite{PRC,IJMPE}.
        The relative energy-shell corrections \cite{MSIS12},
        $\delta E A/\tilde{E}$, are presented too
    by crosses in Figure \ref{fig4}(b).

    The oscillations in Fig.~\ref{fig4}(a) are associated
    with sub-shell effects within
     the major shell, shown in Fig.~\ref{fig4}(b). As seen from
      Fig.~\ref{fig4} and Table I, 
      the results of the MMA2b approach better
          agree with experimental data the larger
      number of states in the low energy states 
      range and the smaller maximal excitation
     energies, $U_{\rm max}$. This is not the case 
     for the MMA1 and other approaches. One of the most pronounced
      cases was considered above for the $^{145}$Nd and $^{142}$Nd nuclei.
     However, sometimes, e.g., for $^{148,150}$Nd,
     the results of
     all approximations are not well distinguished because of almost the same
     $\sigma$. Except for such nuclei,
     for significantly smaller $U_{\rm max}$ the results of the
     MMA2b approach are obviously
         better than the results of
         other approaches.
    Most of the Nd isotopes under consideration are well deformed,
     and the excited
     energy spectra (Ref.~\cite{ENSDFdatabase})
     begin with relatively small energy levels in the low energy states 
     region, except for $^{140,142}$Nd with relatively large $U_{\rm max}$
     (see Table I). 
     Therefore, the pairing
     effects \cite{Ig83,So90,SC19}
     were taken into account in these two nuclei in our Nd calculations
     in the simplest
     version \cite{PRC} of a shift of the excitation energy 
     by pairing condensation energy (see the results which follow the
     asterisks in Table I). 
     Pairing correlations decrease somehow the inverse
     level-density parameter $K$ and smooth
     the sawtooth-like behavior
     of $K(A)$ as function of
     the particle number $A$ (Fig.~\ref{fig3}).
   Accounting for pairing effects, one can observe
     improvement of the results for the same MMA2b approach.
     We should emphasize
     once more that the MMA2 approach 
         for each nucleus is important in the case (ii)
of dominating shell effects (see Subsection~\ref{subsec2-3}).
 As shown in Table I, the isotopic 
  asymmetry effects are more important
      than those of the corresponding one-component nucleon 
case \cite{PRC}. They change significantly
the inverse level-density parameter $K$,
especially for the MMA2b approach.

Thus, for the Nd isotope chain, one can 
see almost one major shell
for mean values of $K(A)$ for each approximation (Fig.~\ref{fig3}).        
  The MMA1 (or SFG and FG) approach yields essentially small values 
  for $K$, which are closer to that of the neutron resonances. 
 Their values differ a little
   from those of the MMA2a approach
   for smaller particle numbers, almost the same for larger particle numbers,
and much smaller than those of the MMA2b approach (Fig.~\ref{fig3}).
As seen clearly from  Figs.~\ref{fig3} and \ref{fig4}, and Table I, 
    in line with results of
Refs.~\cite{ZS16,ZH19}, the obtained values for $K$ 
 within the MMA2 approach can be  
essentially different from those of the MMA1 approach and those 
 of the SFG and FG approaches found, mainly, for the neutron resonances.
 Notice that, as in Refs.~\cite{NPA,PRC,IJMPE}, in all our calculations of
    the statistical level density,
$\rho(E,N,Z)$, we did not use a popular assumption of small spins at large
excitation energies $U$, which is valid for the neutron resonances.
Largely speaking, for the MMA1 approach, one finds values for
$K$ of the same order as
those of the FG and SFG approaches.
These mean values of $K$ are mostly 
close to those of neutron resonances 
in order  of magnitude. The results for $K$ of the
FG and SFG approaches, Eqs.~(\ref{FGtot}) and
(\ref{SPMgennptot}), respectively,
 can be understood because neutron resonances appear 
 relatively at large excitation energies $U$. 
For  these resonances, for the MMA1 approach
    we should not expect 
    such strong shell effects as assumed to be in 
    the MMA2b approach. 
More systematic study of large deformations,  neutron-proton asymmetry, 
and pairing correlations 
 (see Refs.~\cite{Er60,Ig83,So90,AB00,AB03,ZH19,KZ20})
should be 
      taken into account to improve the comparison with experimental data,
      see also preliminary estimates in Ref.~\cite{PRC} for 
  the rare earth and the double magic spherical nucleus $^{208}$Pb.

  \section{Conclusions}
  \label{sec5}

 We have derived the statistical level density $\rho(S)$ as function of the
entropy $S$
within the micro-macroscopic
approximation (MMA) using the mixed micro- and grand-canonical ensembles,
accounting for the neutron-proton asymmetry and collective rotations of nuclei beyond
the saddle point method 
of the standard Fermi gas (FG) model.
This function can be applied for small and, relatively, large entropies
$S$, or excitation energies
$U$ of a nucleus. For a large entropy (excitation energy), one obtains
the 
exponential asymptote of the standard saddle-point
Fermi-gas model, however, with 
significant inverse, $1/S$, power corrections.
For small $S$ one finds the usual finite combinatorics
expansion in  powers of $S^2$. Functionally, the MMA 
linear approximation in the $S^2 \propto U$
    expansion, at small excitation energies $U$,
    coincides with that of the empiric
    constant ``temperature'' model, 
    obtained, however, without using 
    free fitting parameters. Thus,
    the MMA reproduces the 
    well-known Fermi-gas approximation (for large entropy $S$)
    with the 
    constant ``temperature'' model 
    for small entropy $S$,
    also with accounting for the neutron-proton asymmetry and rotational motion. 
The MMA at low excitation energies
    clearly manifests an 
advantage
over the standard full saddle-point 
approaches because of 
no divergences of the MMA in the limit of
small excitation energies, in contrast to all of full saddle-point method,
e.g., the Fermi gas asymptote.
Another advantage takes place 
for nuclei which have
a lot of states in the very low-energy states 
range. In this case, the MMA 
results with only one physical parameter in the least mean-square fit, 
the inverse level-density parameter $K$, 
is usually the better the larger number of the extremely low energy states.
These results are certainly much better than those for the 
Fermi gas model.  
The values of the inverse level-density parameter $K$ 
are compared with those of experimental data for low energy states 
below neutron resonances in 
nuclear spectra of several nuclei. 
The MMA values of $K$ for low energy states can be significantly different
from those of the neutron resonances, studied successfully earlier within
the Fermi gas model.

 We have found a significant shell effects in the MMA level density for the
 nuclear low-energy states range
within the semiclassical periodic-orbit theory. 
In particular, we generalized the known saddle-point method 
results for the level density in
terms of the full shell-structure Fermi gas  (SFG) approximation,
accounting for the shell,
the neutron-proton asymmetry, and rotational effects,
using the periodic-orbit theory. 
Therefore, a reasonable description of the
experimental data for the statistically averaged
level density, obtained by the sample method for low energy states 
was achieved within the MMA with the help of the semiclassical
periodic-orbit theory.
We emphasize the
importance of the shell, neutron-proton asymmetry, and rotational  
effects in these
calculations.   
We obtained 
values of the inverse level-density parameter
$K$ for low-energy states 
range which are essentially different from those of 
 neutron resonances. Taking a long  Nd isotope chain as a typical example,
one finds a saw-toothed behavior of $K(A)$
as function of the particle number $A$ and its 
    remarkable shell oscillation.
We obtained values of $K$ that are significantly 
larger than those obtained for 
neutron resonances, due mainly to accounting for  
the shell effects.
 We show that the semiclassical periodic-orbit theory is helpful in the
low-energy states range for obtaining analytical descriptions of
 the level density
   and energy shell corrections. They are taken into account
   in the linear approximation up to small corrections due to the
   {\it residual} interaction beyond
   the mean field and extended Thomas-Fermi approximation
   within the shell-correction method,
   see Refs.~\cite{BD72,BK72}. The main part of the
    inter-particle interaction is described in terms of 
   the extended Thomas-Fermi counterparts of the
   statistically averaged nuclear potential, and in particular,
   of the level density parameter.

   Our MMA approach for accounting for
       the spin dependence of the level density
   was extended to the collective rotations of deformed nuclei within the
   unified rotation model. The well-known effects of the enhancement due to
   the nuclear collective rotations were considered with accounting for
   the shell structure and neutron-proton asymmetry. This approach might
   be interesting
   in the study of the isomeric states in the strong deformed nuclei
   at high spins due to the shell effects \cite{RB80,St87}.
   We suggest also to work out
   the MMA approach for the description of the collective rotations
   of nuclei, accounting for the phase transitions from the axial to
   non-axial deformations \cite{MM10,MM12}.

 Our approach can be applied to the statistical analysis of
the experimental
data on collective nuclear states, in particular, for the nearest-neighbor
spacing distribution calculations within the Wigner-Dyson theory of
quantum chaos \cite{Ze96,Ze16,GK11}.
As the semiclassical periodic-orbit 
MMA is the better the larger particle number
    in a Fermi system, one can apply this method 
    also for study of the metallic clusters and quantum dots
    in terms of the statistical level density, and  of several problems in
    nuclear astrophysics.
As perspectives,
the 
collective rotational excitations at
large nuclear angular momenta and
deformations, 
as well as more consequently 
pairing correlations, all with a more systematic accounting
for
the neutron-proton asymmetry, will be 
taken into account in a future  work.
    In this way, we expect to improve the comparison of the
theoretical evaluations
with
experimental data on the level density parameter 
significantly for energy levels below the neutron resonances.

\section*{Acknowledgments}

The authors gratefully acknowledge 
D.~Bucurescu, R.K.~Bhaduri, M.~Brack,
 A.N.~Gorbachenko, and V.A.~Plujko
for creative discussions.
A.G.M. would like to thank  the Cyclotron 
Institute of Texas A\&M University for the nice hospitality extended to him.
 This work was supported in part by the  budget program
"Support for the development
of priority areas of scientific researches",  the project of the
Academy of Sciences of Ukraine (Code 6541230, no. 0122U000848).
S.S. and A.G.M. are partially supported by the US Department of Energy
under Grant no. DE-FG03-93ER-40773.

\appendix

\renewcommand{\theequation}{A\arabic{equation}}
\renewcommand{\thesubsection}{A\arabic{subsection}}
\setcounter{equation}{0}

\section{Semiclassical periodic-orbit theory 
  for isotopically asymmetric rotating system}\label{appA}

Introducing the isotopic index $\tau=\{n,p\}$ for isotopically asymmetric nuclear system, one
can present the partition function $\ln\mathcal{Z}$ as sum of $\ln\mathcal{Z}_\tau$,
$\ln\mathcal{Z}=\ln\mathcal{Z}_n+\ln\mathcal{Z}_p$.
In the
case of the ``parallel'' rotation (alignment of the angular momenta of
individual particles along the symmetry axis $0z$),
one has for a spherical and 
axial symmetric  potential
the explicit 
$\tau$ partition-function component: 
\bea\l{parfun}
&\ln \mathcal{Z}_\tau= 
\sum\limits_{i}\ln\left\{1 +
\exp\left[\beta\left(\lambda_\tau - 
    \varepsilon_i+\hbar \omega m_i\right)\right]\right\}\nonumber\\
&\approx \int\limits_0^{\infty}\d \varepsilon
\int\limits_{-\infty}^{\infty}\mbox{d} m~g_\tau(\varepsilon,m)
\ln\left\{1+\right.\nonumber\\
&+\left.\exp\left[\beta\left(\lambda -
    \varepsilon+\hbar \omega m\right)\right]\right\}~.
\eea
Here, $\varepsilon_i$ and $m_i$ are the single-particle
(s.p.) energies and projections of the angular momentum on the
symmetry
axis $Oz$ of the quantum states $i$ of the $\tau$ system
in the axially symmetric
mean-field potential well, respectively.
In the transformation
from the sum to an
integral, we introduced  the s.p. level density $g_\tau(\varepsilon,m)$ as a sum of
the smooth, $\tilde{g}_\tau$, and oscillating shell, $\delta g_\tau$, 
components; see Eq.~(\ref{gdecomp}).
The Strutinsky smoothed 
level-density component $\tilde{g}_\tau$
can be well approximated by
the ETF level density $g^{(\tau)}_{\rm \tt{ETF}}$, $\tilde{g}_\tau\approx g^{(\tau)}_{\rm \tt{ETF}}$.
For the spherical case, as an example, the 
level density in the TF
approximation, $g^{(\tau)}_{\rm \tt{TF}}$, for
any fixed $\tau$ is given by \cite{Be48}
\be\l{tildeg}
\tilde{g}\approx g^{}_{\rm \tt{TF}}=\frac{\mu d_s}{\pi\hbar}\int\limits_{|m|}^{\ell_0^{}}\d \ell
\int\limits_{r_{\rm min}}^{r_{\rm max}}
\d r \left[2 \mu\left(\varepsilon - V(r)\right)-\hbar^2l^2/r^2\right]^{-1}~,
\ee
where $\mu$ is the nucleon mass, $d_s$ is the spin (spin-isospin) degeneracy,
$\ell_0$ is the maximum of a  possible
angular momentum of 
nucleon with energy  $\varepsilon$
in a spherical potential well $V(r)$, and $~r_{\rm min}$
and $r_{\rm max}$ are the turning points. We assume that the
asymmetry parameter $X^2=[(N-Z)/A]$ is
small and $\lambda_n\approx \lambda_p\approx \lambda$.
    Therefore, the fixed sub(super)script
    $\tau$ is omitted here and below when it will not lead
    to a misunderstanding.
For the oscillating component
$\delta g_{\rm scl}(\varepsilon,m)$ of the 
level density $g(\varepsilon,m)$
[Eq.~(\ref{gdecomp})]
we use, in the spherical case 
(at a given $\tau$),
the following
semiclassical expression \cite{KM79}
derived in Ref.~\cite{MK78}:
\be\l{goscemsph}
\delta g_{\rm scl}(\varepsilon,m)=\sum^{}_{\rm PO}\frac{1}{2 \ell_{\rm PO}}
\theta\left(\ell_{\rm PO}-|m|\right)~g^{}_{\rm PO}(\varepsilon)~.
\ee
The sum in Eq.~(\ref{goscemsph})
is taken over the classical periodic orbits (PO) with 
angular momenta $\ell_{\rm PO}\geq |m|$. In 
the sum of Eq.~(\ref{goscemsph}),
$g^{}_{\rm PO}(\varepsilon)$ is the partial
contribution of the PO to the oscillating
    part $\delta g_{\rm scl}(\varepsilon)$
of the total semiclassical 
level density $g_{\rm scl}(\varepsilon)$ (without limitations
on the projection
$m$ of the particle angular momentum)
with 
\be\l{goscsem}
\delta g_{\rm scl}(\varepsilon)=\sum^{}_{\rm PO}g^{}_{\rm PO}(\varepsilon)~,
\ee
where
\be\l{goscPO}
g^{}_{\rm PO}(\varepsilon)=\mathcal{A}_{\rm PO}(\varepsilon)
~\cos\left[\frac{1}{\hbar}\mathcal{S}_{\rm PO}(\varepsilon)-
\frac{\pi}{2} \mathcal{M}^{}_{\rm PO}
-\phi^{}_0\right].
\ee
 Here, $\mathcal{S}_{\rm PO}(\varepsilon)$ is the classical action along the
 PO, $\mathcal{M}^{}_{\rm PO}$ is the so called Maslov index determined by
the catastrophe points (turning and caustic points) along the PO, and
$\phi^{}_0$ is an additional shift of the phase coming from the dimension
of the problem and degeneracy of the POs. The amplitude
$\mathcal{A}_{\rm PO}(\varepsilon)$  in Eq.~(\ref{goscPO}) is 
a smooth function of
the energy
$\varepsilon$,  depending 
on the PO stability factors
\cite{SM76,BB03,MY11}.
      For a spherical cavity one has the famous explicitly analytical
 Balian-Bloch formula \cite{BB03,SM76}. The Gaussian
       local averaging of the level density shell correction 
          $\delta g^{}_{\rm scl}(\varepsilon)$ [Eq.~(\ref{goscsem})]
          over the quasiparticle 
          energy spectrum
      $\varepsilon_i$ near the Fermi surface $\varepsilon^{}_F$ can be done
  analytically by using the linear expansion of
  relatively 
  smooth PO action
      integral $\mathcal{S}_{\rm PO}(\varepsilon)$ 
          near $\varepsilon^{}_F$ as function of $\varepsilon$
 with a Gaussian width parameter $\Gamma$ \cite{SM76,BB03,MY11},
\be\l{avden}
\delta g^{(\Gamma)}_{\rm scl}(\varepsilon) \cong
\sum^{}_{\rm PO}g^{}_{\rm PO}(\varepsilon)~
\exp\left[-\left(\frac{\Gamma t^{}_{\rm PO}}{2\hbar}\right)^2\right]~,
\ee
where $t^{}_{\rm PO}=\partial S_{\rm PO}/\partial \varepsilon$ is the period
of particle motion along the PO.
    All the expressions presented above, 
    except for
    Eqs.~(\ref{tildeg}) and (\ref{goscemsph}), can be applied for the axially-symmetric
    potentials, e.g., for the spheroidal cavity
    \cite{SM77,MA02,MY11} and deformed
    harmonic oscillator
    \cite{Ma78,BB03}.
    For the smooth part of the level density, $\tilde{g}$, and corresponding
    nuclear energy,
$\tilde{E}$ (see Ref.~\cite{BD72} for the SCM) within the POT, we
use the semiclassical extended Thomas-Fermi
    approximations,  $g^{}_{\rm ETF}$ and 
    $ E_{\rm ETF}$, respectively. These
 expressions are well derived and explained in Refs.~\cite{BG85,BB03,KS20}. 
 The smooth  chemical potential $\tilde{\lambda}$ in the SCM is 
 the root of equations
  $ \mathcal{N}_\tau=
       \int_{0}^{\tilde{\lambda}}\mbox{d} \varepsilon~\tilde{g}_\tau(\varepsilon)$, and
       $\lambda \approx \tilde{\lambda}$ in the POT.
 The chemical potential $\lambda$ (or  $\tilde{\lambda}$)
 is approximately the solution of the corresponding
 particle number conservation equation:
  \be\l{chempoteq}
    \mathcal{N}_\tau~=\int_{0}^{\lambda} \mbox{d} \varepsilon~g_\tau(\varepsilon)~.
\ee
The smooth quantity $\tilde{\Theta}\approx\Theta^{}_{\rm \tt{ETF}}$ in
Eq.~(\ref{TFpotF}) is
   the ETF (rigid-body) moment of inertia for the statistical
   equilibrium rotation,
\bea\l{rigMIpar}
& \Theta^{}_{\tt{ETF}}=
  \mu\int \d {\bf r}~\rho_{\tt{ETF}}({\bf r})~(x^2+y^2)\nonumber\\
&\approx \hbar^2
\langle \widetilde{m^2}\rangle~g_{\tt{ETF}}\left(\lambda\right)~,
\eea
where $\rho^{}_{\tt{ETF}}({\bf r})$
is the ETF
particle number density. 
For a ``parallel'' rotation, 
$\langle \widetilde{m^2}\rangle$
is 
the smooth component of
the square
of the angular momentum
projection, $\langle m^2\rangle$, of nucleon.
Here and below we neglect a small change 
in the chemical
 potential $\lambda$, 
due to the internal nuclear  thermal and rotational excitations, 
which can be approximated by the Fermi energy $\varepsilon^{}_F$,
$\lambda\approx \varepsilon^{}_F$.

 The oscillating semiclassical component
$\delta \Omega\left(\beta,\lambda,\omega\right)$ 
of the
sum (\ref{OmFnp}) 
corresponds to the oscillating part
$\delta g_{\rm scl}(\varepsilon,m)$ of the
level density $g_{\rm scl}(\varepsilon,m)$ [Eq.~(\ref{gdecomp})]; 
see, e.g.,
  Eq.~(\ref{goscemsph}) for the spherical case and Refs.~\cite{SM76,KM79,MK78}.
In expanding the action $\mathcal{S}_{\rm PO}(\varepsilon)$ as function of
the energy $\varepsilon$ near the chemical potential $\lambda$ in
powers of $\varepsilon-\lambda$ up to linear term, 
one can use Eqs.~(\ref{goscsem}) and (\ref{goscPO}); 
 see also Eqs.~(\ref{FESCFnp}), (\ref{dFESCFnp}), and
(\ref{dEPO0Fnp}).
Then, integrating by parts, one obtains 
from Eqs.~(\ref{parfun}),
(\ref{OmFnp}), and (\ref{TFpotF})
the shell correction $\delta\Omega$ in 
the adiabatic approximation,
$\hbar \ell^{2}_F\omega 
\ll \lambda$,  
where $\hbar \ell^{}_F$ is the maximal 
particle spin at the Fermi surface. For the spherical case, one finds its simple
explicit result:
\bea\l{potoscpar}
&\delta \Omega \cong \delta \Omega_{\rm scl}\left(\beta,\lambda,\omega\right)
=\delta F_{\rm scl}\left(\beta,\lambda,\omega\right)\nonumber\\
&=\delta F_{\rm scl}\left(\beta,\lambda\right)
-\frac{1}{2}\delta \Theta~\omega^2,
\eea
where $\delta F_{\rm scl}\left(\beta,\lambda\right)$ is the
semiclassical free-energy shell correction of nonrotating nucleus
($\omega=0$);
see Eqs.~(\ref{FESCFnp})
and (\ref{dFESCFnp}). For the spherical mean-field approach,
the shell correction $\delta\Theta$
to the moment of inertia $\Theta$ [Eq.~(\ref{MI})]
can be presented as
\be\l{dMIsph}
\delta \Theta \cong \delta \Theta_{\rm scl} =
\frac13\sum^{}_{\rm PO}t^{2}_{\rm PO}l^{2}_{\rm PO}~F_{\rm PO}~.
\ee
 In deriving the expressions for the
free energy shell correction,
$\delta F_{\rm scl}$, 
and the potential, $\delta \Omega_{\rm scl}$, the action
$\mathcal{S}_{\rm PO}(\varepsilon)$ in their integral representations over
$\varepsilon $ with the semiclassical level-density shell correction,
$\delta g(\varepsilon)$, Eqs.~(\ref{goscsem}) and (\ref{goscPO}),
was expanded 
near the chemical potential $\lambda$ 
    up to the second order corrections over $\varepsilon-\lambda$.
 Then, we integrated by parts over $\varepsilon$, as in
the semiclassical calculations of the energy shell correction,
$\delta E_{\rm scl}$  \cite{SM76,BB03}. 
We used the expansion of
$\delta \Omega(\beta,\lambda,\omega)$ over a relatively small
rotation frequency $\omega$,
$\hbar \ell^{2}_F\omega/\lambda\ll 1$, up to
quadratic terms. In the adiabatic approximation, one
can simplify the decomposition of the potential
$\Omega$ 
[Eq.~(\ref{OmFnp}) with Eq.~(\ref{gdecomp})] in
terms of smooth and oscillating POT components,
    Eqs.~(\ref{TFpotF}) and
    (\ref{potoscparFnp}), or (\ref{potoscpar}) for a given isotopic value of
    $\tau$,
%
\be\l{Omad}
\Omega \approx
E_0-\frac{a}{\beta^2}-\lambda A-\frac12~\Theta~\omega^2~;
\ee
see also Eq.~(\ref{MI}) for the moment of inertia $\Theta$.
The level density parameter
$a$ is given by Eq.\ (\ref{denparnp}) modified, however, by the rotational
$\omega^2$ corrections:
\be\l{a0par}
a \approx \frac{\pi^2}{6}\left[g\left(\lambda\right)
+\frac{\omega^2}{6} \sum^{}_{\rm PO}g^{}_{\rm PO}\left(\lambda\right)
~t^{2}_{\rm PO}~l^{2}_{\rm PO}\right]~.
\ee
 The second term in the square brackets is  explicitly
    presented for the spherical potential.
Equation (\ref{Omad}), which is  valid for arbitrary axially-symmetric potential,
contains shell effects through the ground-state energy $E^{}_0$,
the level density parameter $a$, Eq.~(\ref{a0par}),
and moment of inertia
(MI), Eqs.~(\ref{MI}) and 
(\ref{dMIsph}).
 Non-adiabatic effects for large $\omega$,
    considered in 
Ref.~\cite{KM79} for the spherical case, are out
of the scope of this work.
In Eq.~(\ref{potoscpar}), the period of motion along a PO,
$t^{}_{\rm PO}(\varepsilon)=\partial S_{\rm PO}(\varepsilon)/\partial \varepsilon$,
and
the PO angular momentum of particle, $\ell^{}_{\rm PO}(\varepsilon)$,
are taken at $\varepsilon=\lambda$. 
For large excitation energies,
$\beta=\beta^{\ast}=1/T$ ($T$ is the temperature), 
one arrives from Eqs.~(\ref{FESCFnp}), (\ref{dFESCFnp}), and (\ref{potoscpar}) at
the well-known expression for the semiclassical free-energy shell correction
of the POT \cite{KM79,BB03}, $\delta F=\delta \Omega$
    (in their specific variables).
These shell corrections decrease exponentially with
increasing temperature $T$. For the opposite limit to the yrast line
(zero excitation energy $U$, $\beta^{-1}\sim T \rightarrow 0$), one obtains from
$\delta \Omega$, Eq.~(\ref{potoscpar}), the well-known
POT approximation \cite{SM76,BB03} to the energy shell correction
$\delta E$, modified, however, by
the frequency $\omega $ dependence.

The POT shell effect component of the free energy, $\delta F_{\rm scl}$ [Eqs.~(\ref{FESCFnp}),
and (\ref{dFESCFnp})],
is related in the nonthermal and nonrotational
limit to the energy
    shell correction of a cold nucleus,
$\delta E_{\rm scl}$
\cite{SM76,BB03,MY11,MK16},
\be\l{escscl}
\delta E_{\rm scl} = \sum_{\rm PO}E_{\rm PO}= \sum_{\rm PO}\frac{\hbar^2}{t_{\rm PO}^2}\,
g^{}_{\rm PO}(\lambda) 
~,
\ee
 where $E_{\rm PO}$ is the partial PO component [Eq.~(\ref{dEPO0Fnp})] of
the energy shell correction
$\delta E$. 
Within the POT, $\delta E_{\rm scl}$
is determined, in turn, by 
the oscillating level density $\delta g_{\rm scl}(\lambda)$, 
see Eqs.~(\ref{goscsem}) and (\ref{goscPO}).

The chemical potential $\lambda $, 
    for a fixed isotopic value of $\tau$,
can be approximated by the Fermi energy
$\vareps^{}_F$, up to small excitation-energy and rotation frequency
corrections ($T\ll \lambda$
for the saddle point
value $T=1/\beta^\ast$ if exists, and
$\hbar\ell^{}_F\omega/\lambda \ll 1$).
  It is determined by the particle-number conservation condition,
 Eq.~(\ref{Seqsdnp}),  which can be written in a simple form (\ref{chempoteq}) 
      with the total POT level density
  $g(\vareps)\cong g^{}_{\rm scl}=g^{}_{\rm \tt{ETF}} +\delta g^{}_{\rm scl}$,
as a good approximation to the integrand of the particle number
conservation equation (\ref{Seqsdnp}) for $\lambda_n$ and $\lambda_p$
 and a given $\tau$. One now needs
to solve Eq.~(\ref{chempoteq}) for a given particle number, $\mathcal{N}_\tau$,
to determine their chemical potential $\lambda_\tau$ as function of
$\mathcal{N}_\tau$. To solve this equation with good accuracy, it is helpful to
use the expression for the integrand which is equal to the level density of
the shell correction method \cite{BD72}, see also Refs.~\cite{PRC,IJMPE}.
The mean chemical potential $\lambda$ ($\lambda_n\approx\lambda_p\approx\lambda$)
is needed in Eq.~(\ref{escscl}) to obtain the semiclassical energy shell
corrections $\delta E_{\rm scl}$.

For a major (neutron or proton)
shell structure near the Fermi energy surface,
$\varepsilon\approx \lambda$,
the POT shell 
 correction, $\delta E_{\rm scl}$ [Eq.~(\ref{escscl})]
 is in
 fact approximately proportional to that of 
$\delta g_{\rm scl}(\lambda)$ 
[Eqs.\ (\ref{goscsem}) and (\ref{goscPO})].
Indeed, the rapid convergence of the PO sum in Eq.~(\ref{escscl})
is guaranteed by the 
factor in front of the density component $g^{}_{\rm PO}$,
Eq.\ (\ref{goscPO}), a factor 
which is inversely proportional to the square of the period
time $t^{}_{\rm PO}(\lambda)$ 
along 
the PO. Therefore, only POs with 
short periods which occupy a 
significant 
phase-space volume near the Fermi surface will contribute.
These orbits are responsible for the
major shell structure, that is related to a Gaussian averaging width,
$\Gamma\approx \Gamma_{\rm sh}$, which is much larger
than the distance between neighboring s.p. states but much smaller
than the distance
$\mathcal{D}_{\rm sh} $ between major shells near the Fermi surface.
According to the POT \cite{SM76,BB03,MY11},
the distance between major shells, $\mathcal{D}_{\rm sh}$, is
determined by a
mean period of the  
shortest and most degenerate POs, $\langle t^{}_{\rm PO}\rangle$,
for $\lambda_\tau\approx \lambda$
\cite{SM76,BB03}:
\be\l{periode}
\mathcal{D}^{}_{\rm sh} \cong 
\frac{2\pi \hbar}{\langle t^{}_{\rm PO}\rangle} 
\approx \frac{\lambda}{A^{1/3}}~,
\ee
where $A=N+Z$.
Taking the factor in front of
$g^{}_{\rm PO}$ in 
the energy shell correction
$\delta E^{}_{\rm scl}$, Eq.~(\ref{escscl}), 
off the sum over the POs, one arrives at
Eq.~(\ref{dedgnp}) for the semiclassical energy-shell correction
\cite{SM76,SM77,MY11,MK16}. Differentiating Eq.~(\ref{escscl}) 
 using (\ref{goscPO}) with respect to $\lambda$
and keeping only the dominating terms coming from 
differentiation of the sine of the action phase argument,
$S/\hbar \sim A^{1/3}$, one finds the useful relationship: 
\be\l{d2Edl2}
\frac{\partial^2\delta E^{}_{\rm PO}}{\partial\lambda^2}\approx -
\delta g^{}_{\rm PO}~.
\ee
By the same semiclassical arguments, the dominating contribution to
the double derivative $g''(\lambda)$
for a major shell structure is given by
\be\l{d2g}
\frac{\partial^2 g}{\partial\lambda^2}\approx
\sum^{}_{\rm PO}\frac{\partial^2\delta g^{}_{\rm PO}}{\partial\lambda^2}
  \approx -\left(\frac{2\pi}{\mathcal{D}^{}_{\rm sh}}\right)^2 \delta g(\lambda)~.
\ee
Again, as in the derivation of 
 Eqs.~(\ref{dedgnp}) and (\ref{d2Edl2}),
 for a major shell
structure, we take the averaged smooth characteristics for the main
shortest POs which occupy the largest phase-space volume off the PO sum.  

\renewcommand{\theequation}{B\arabic{equation}}
\renewcommand{\thesubsection}{B\arabic{subsection}}
\setcounter{equation}{0}
  
\section{
Full SPM for a shell structure Fermi gas (SFG) asymptote}
\l{appB}

Taking 
the integral (\ref{rhoE1Fnp}) over $\beta $ by the standard
SPM,
one can expand, up to second order terms, the exponent argument $S(\beta)=\beta U+a/\beta$
near the 
 saddle point $\beta=\beta^\ast$,
\be\l{expbeta}
S(\beta)=\beta^\ast U+a/\beta^\ast + \frac12
\left(\frac{2a}{\beta^3}\right)^\ast(\beta-\beta^\ast)^2.
\ee
The first derivative disappears because of the SPM condition: 
\be\l{spmcondbeta}
\left(\frac{\partial S}{\partial \beta}\right)^\ast\equiv U-
\frac{a}{(\beta^{\ast})^2}=0~,
\ee
from which one finds the standard expression for the excitation energy $U$
through the 
 saddle point $\beta^\ast=1/T$, i.e., $U=aT^2$.
Taking the pre-exponential Jacobian multiplier off the integral
 over $\beta$ in Eq.~(\ref{rhoE1Fnp}) 
we substitute Eq.~(\ref{expbeta}) for $S(\beta) $
into Eq.~(\ref{rhoE1Fnp}). 
Changing the integration variable $\beta$ to the new variable $z$,
$z^2=(-\partial^2 S/\partial \beta^2)^{\ast}(\beta-\beta^\ast)^2/2$,
and then, calculating the error integral over $z$ by extending the
integration range to infinity, one obtains Eq.~(\ref{SPMgennp}).
Here we used a general expression (\ref{Jacnp}) for the Jacobian
at the saddle point 
condition (\ref{expbeta}) for $\beta=\beta^\ast$. 
The critical quantity for these derivations is the ratio $\xi^\ast$,
see 
Eq.~(\ref{xiparnp}) for $\xi$ taken at $\beta=\beta^\ast$, $\xi=\xi^\ast$,
which is approximately proportional
to the semiclassical POT 
  energy shell correction, Eq.~(\ref{dedgnp}) (see Appendix \ref{appA}).

\end{document}